\documentstyle[10pt,epsf,epsfig,dp_delphititle]{dp_delphi}
\makeindex
\def\DpPaperGroup{EP}
\def\DpPaperRef{99-174}
\def\DpDate{16 December 1999}
\def\DpAuthors{DELPHI Collaboration}
\def\DpSubmit{(Phys. Lett. B475(2000)407)}
\def\DpTitle{{Measurement of the \boldmath $\bar{\mathrm B} \rightarrow 
{\mathrm D}^{(*)} \pi \ell \bar{\nu}_{\ell}$ \\ 
Branching Fraction}}

\newcommand{\epem} {\ifmmode{e^+e^-}\else{$e^+e^-$}\fi}
\newcommand{\qqbar} {\ifmmode{q\bar{q}}\else{$q\bar{q}$}\fi}

\newcommand{\eps}{{\ifmmode \varepsilon \else $\varepsilon$\fi}}
\newcommand{\thetac}{{\ifmmode \theta_C\else $\theta_C$\fi}}

\newcommand{\BC}{\begin{center}}
\newcommand{\EC}{\end{center}}
\newcommand{\BE}{\begin{equation}}
\newcommand{\EE}{\end{equation}}
\newcommand{\BEA}{\begin{eqnarray}}
\newcommand{\EEA}{\end{eqnarray}}
\newcommand{\BA}{\begin{array}}
\newcommand{\EA}{\end{array}}
\newcommand{\BI}{\begin{itemize}}
\newcommand{\EI}{\end{itemize}}
\newcommand{\BF}{\begin{figure}}
\newcommand{\EF}{\end{figure}}
\newcommand{\BT}{\begin{table}}
\newcommand{\ET}{\end{table}}
\newcommand{\BTB}{\begin{tabular}}
\newcommand{\ETB}{\end{tabular}}
\newcommand\BM{\begin{minipage}}
\newcommand\EM{\end{minipage}}

\newcommand{\Erg}[3]{\ifmmode{#1\pm#2_{stat.}\pm#3_{sys.}}\else{$#1\pm#2_{stat.}\pm#3_{sys.}$}\fi}
\newcommand{\erg}[3]{\ifmmode{\scriptstyle#1\,\pm\,#2\,\pm\,#3}\else{$\scriptstyle#1\,\pm\,#2\,\pm\,#3$}\fi}
\begin{document}
\makeatletter
\newcount\@tempcntc
\def\@citex[#1]#2{\if@filesw\immediate\write\@auxout{\string\citation{#2}}\fi
  \@tempcnta\z@\@tempcntb\m@ne\def\@citea{}\@cite{\@for\@citeb:=#2\do
    {\@ifundefined
       {b@\@citeb}{\@citeo\@tempcntb\m@ne\@citea\def\@citea{,}{\bf ?}\@warning
       {Citation `\@citeb' on page \thepage \space undefined}}%
    {\setbox\z@\hbox{\global\@tempcntc0\csname b@\@citeb\endcsname\relax}%
     \ifnum\@tempcntc=\z@ \@citeo\@tempcntb\m@ne
       \@citea\def\@citea{,}\hbox{\csname b@\@citeb\endcsname}%
     \else
      \advance\@tempcntb\@ne
      \ifnum\@tempcntb=\@tempcntc
      \else\advance\@tempcntb\m@ne\@citeo
      \@tempcnta\@tempcntc\@tempcntb\@tempcntc\fi\fi}}\@citeo}{#1}}
\def\@citeo{\ifnum\@tempcnta>\@tempcntb\else\@citea\def\@citea{,}%
  \ifnum\@tempcnta=\@tempcntb\the\@tempcnta\else
   {\advance\@tempcnta\@ne\ifnum\@tempcnta=\@tempcntb \else \def\@citea{--}\fi
    \advance\@tempcnta\m@ne\the\@tempcnta\@citea\the\@tempcntb}\fi\fi}
 
\makeatother
\begin{titlepage}
\pagenumbering{roman}
\CERNpreprint{\DpPaperGroup}{\DpPaperRef} 
\date{{\small\DpDate}} 
\title{\DpTitle} 
\address{\DpAuthors} 
\begin{shortabs} 
\noindent
%
\noindent
A study of $b$ semileptonic decays into $D$, $D\pi^{\pm}$ and $D^*\pi^{\pm}$ final states is presented.
The $D^0$, $D^+$ and $D^{*+}$ mesons are exclusively reconstructed in $Z$ decay data recorded from 1992 to 1995
in the DELPHI experiment at LEP.
The overall branching fractions are measured to be:
\begin{eqnarray*}
{\mathrm BR}(b \rightarrow D^0 \ell^- \bar{\nu}_{\ell} X)    &=& 
 (7.04\pm0.34\;(stat)\pm0.36\;(syst.exp)\pm0.17\;({\mathrm BR}_D))\% \\
{\mathrm BR}(b \rightarrow D^+ \ell^- \bar{\nu}_{\ell} X)    &=& 
 (2.72\pm0.19\;(stat)\pm0.16\;(syst.exp)\pm0.18\;({\mathrm BR}_D))\% \\
{\mathrm BR}(b \rightarrow D^{*+} \ell^- \bar{\nu}_{\ell} X) &=& 
 (2.75\pm0.17\;(stat)\pm0.13\;(syst.exp)\pm0.09\;({\mathrm BR}_D))\% 
\end{eqnarray*}
where the $D^0$ and $D^+$ results include also contributions from 
$D^{*0}$ and $D^{*+}$ decays.
A fit to the distribution of the $\pi^{\pm}$ impact parameter
to the primary interaction vertex
provides a measurement of the $b$ semileptonic branching fractions
into the $D^0 \pi^{\pm} X$, $D^+ \pi^{\pm} X$ and $D^{*+} \pi^{\pm} X$ final states.
Assuming that single pion decay modes of $B$ mesons dominate, the partial rates
for $\bar{B} \rightarrow D \pi \ell^- \bar{\nu_{\ell}}$
and $\bar{B} \rightarrow D^{*} \pi \ell^- \bar{\nu_{\ell}}$
have been obtained, corresponding to a total branching fraction:
\begin{eqnarray*}
{\mathrm BR}(\bar{B} \rightarrow D \pi \ell^- \bar{\nu_{\ell}}) + 
{\mathrm BR}(\bar{B} \rightarrow D^{*} \pi \ell^- \bar{\nu_{\ell}}) 
                      = (3.40\pm0.52\;(stat)\pm0.32\;(syst))\%  \; .
\end{eqnarray*}
This result agrees well with the observed difference between
the total $B$ semileptonic branching fraction and the sum
of the $\bar{B} \rightarrow D \ell^- \bar{\nu_{\ell}}$
and $D^* \ell^- \bar{\nu_{\ell}}$ branching fractions.

\end{shortabs}
\vfill
\begin{center}
\DpSubmit \ \\ 
\end{center}
\vfill
\clearpage
\headsep 10.0pt
\addtolength{\textheight}{10mm}
\addtolength{\footskip}{-5mm}
\begingroup
%
\newcommand{\DpName}[2]{\hbox{#1$^{\ref{#2}}$},\hfill}
\newcommand{\DpNameTwo}[3]{\hbox{#1$^{\ref{#2},\ref{#3}}$},\hfill}
\newcommand{\DpNameThree}[4]{\hbox{#1$^{\ref{#2},\ref{#3},\ref{#4}}$},\hfill}
\newskip\Bigfill \Bigfill = 0pt plus 1000fill
\newcommand{\DpNameLast}[2]{\hbox{#1$^{\ref{#2}}$}\hspace{\Bigfill}}
%
\footnotesize
\noindent
\DpName{P.Abreu}{LIP}
\DpName{W.Adam}{VIENNA}
\DpName{T.Adye}{RAL}
\DpName{P.Adzic}{DEMOKRITOS}
\DpName{I.Ajinenko}{SERPUKHOV}
\DpName{Z.Albrecht}{KARLSRUHE}
\DpName{T.Alderweireld}{AIM}
\DpName{G.D.Alekseev}{JINR}
\DpName{R.Alemany}{VALENCIA}
\DpName{T.Allmendinger}{KARLSRUHE}
% ...
\normalsize
\endgroup
\titlefoot{Department of Physics and Astronomy, Iowa State
     University, Ames IA 50011-3160, USA
    \label{AMES}}
\titlefoot{Physics Department, Univ. Instelling Antwerpen,
     Universiteitsplein 1, B-2610 Antwerpen, Belgium \\
     \indent~~and IIHE, ULB-VUB,
     Pleinlaan 2, B-1050 Brussels, Belgium \\
     \indent~~and Facult\'e des Sciences,
     Univ. de l'Etat Mons, Av. Maistriau 19, B-7000 Mons, Belgium
    \label{AIM}}
\titlefoot{Physics Laboratory, University of Athens, Solonos Str.
     104, GR-10680 Athens, Greece
    \label{ATHENS}}
\titlefoot{Department of Physics, University of Bergen,
     All\'egaten 55, NO-5007 Bergen, Norway
    \label{BERGEN}}
\titlefoot{Dipartimento di Fisica, Universit\`a di Bologna and INFN,
     Via Irnerio 46, IT-40126 Bologna, Italy
    \label{BOLOGNA}}
\titlefoot{Centro Brasileiro de Pesquisas F\'{\i}sicas, rua Xavier Sigaud 150,
     BR-22290 Rio de Janeiro, Brazil \\
     \indent~~and Depto. de F\'{\i}sica, Pont. Univ. Cat\'olica,
     C.P. 38071 BR-22453 Rio de Janeiro, Brazil \\
     \indent~~and Inst. de F\'{\i}sica, Univ. Estadual do Rio de Janeiro,
     rua S\~{a}o Francisco Xavier 524, Rio de Janeiro, Brazil
    \label{BRASIL}}
\titlefoot{Comenius University, Faculty of Mathematics and Physics,
     Mlynska Dolina, SK-84215 Bratislava, Slovakia
    \label{BRATISLAVA}}
\titlefoot{Coll\`ege de France, Lab. de Physique Corpusculaire, IN2P3-CNRS,
     FR-75231 Paris Cedex 05, France
    \label{CDF}}
\titlefoot{CERN, CH-1211 Geneva 23, Switzerland
    \label{CERN}}
\titlefoot{Institut de Recherches Subatomiques, IN2P3 - CNRS/ULP - BP20,
     FR-67037 Strasbourg Cedex, France
    \label{CRN}}
\titlefoot{Now at DESY-Zeuthen, Platanenallee 6, D-15735 Zeuthen, Germany
    \label{DESY}}
\titlefoot{Institute of Nuclear Physics, N.C.S.R. Demokritos,
     P.O. Box 60228, GR-15310 Athens, Greece
    \label{DEMOKRITOS}}
\titlefoot{FZU, Inst. of Phys. of the C.A.S. High Energy Physics Division,
     Na Slovance 2, CZ-180 40, Praha 8, Czech Republic
    \label{FZU}}
\titlefoot{Dipartimento di Fisica, Universit\`a di Genova and INFN,
     Via Dodecaneso 33, IT-16146 Genova, Italy
    \label{GENOVA}}
\titlefoot{Institut des Sciences Nucl\'eaires, IN2P3-CNRS, Universit\'e
     de Grenoble 1, FR-38026 Grenoble Cedex, France
    \label{GRENOBLE}}
\titlefoot{Helsinki Institute of Physics, HIP,
     P.O. Box 9, FI-00014 Helsinki, Finland
    \label{HELSINKI}}
\titlefoot{Joint Institute for Nuclear Research, Dubna, Head Post
     Office, P.O. Box 79, RU-101 000 Moscow, Russian Federation
    \label{JINR}}
\titlefoot{Institut f\"ur Experimentelle Kernphysik,
     Universit\"at Karlsruhe, Postfach 6980, DE-76128 Karlsruhe,
     Germany
    \label{KARLSRUHE}}
\titlefoot{Institute of Nuclear Physics and University of Mining and Metalurgy,
     Ul. Kawiory 26a, PL-30055 Krakow, Poland
    \label{KRAKOW}}
\titlefoot{Universit\'e de Paris-Sud, Lab. de l'Acc\'el\'erateur
     Lin\'eaire, IN2P3-CNRS, B\^{a}t. 200, FR-91405 Orsay Cedex, France
    \label{LAL}}
\titlefoot{School of Physics and Chemistry, University of Lancaster,
     Lancaster LA1 4YB, UK
    \label{LANCASTER}}
\titlefoot{LIP, IST, FCUL - Av. Elias Garcia, 14-$1^{o}$,
     PT-1000 Lisboa Codex, Portugal
    \label{LIP}}
\titlefoot{Department of Physics, University of Liverpool, P.O.
     Box 147, Liverpool L69 3BX, UK
    \label{LIVERPOOL}}
\titlefoot{LPNHE, IN2P3-CNRS, Univ.~Paris VI et VII, Tour 33 (RdC),
     4 place Jussieu, FR-75252 Paris Cedex 05, France
    \label{LPNHE}}
\titlefoot{Department of Physics, University of Lund,
     S\"olvegatan 14, SE-223 63 Lund, Sweden
    \label{LUND}}
\titlefoot{Universit\'e Claude Bernard de Lyon, IPNL, IN2P3-CNRS,
     FR-69622 Villeurbanne Cedex, France
    \label{LYON}}
\titlefoot{Univ. d'Aix - Marseille II - CPP, IN2P3-CNRS,
     FR-13288 Marseille Cedex 09, France
    \label{MARSEILLE}}
\titlefoot{Dipartimento di Fisica, Universit\`a di Milano and INFN,
     Via Celoria 16, IT-20133 Milan, Italy
    \label{MILANO}}
\titlefoot{Universit\`a degli Studi di Milano - Bicocca,
     Via Emanuelli 15, IT-20126 Milan, Italy
    \label{MILANO2}}
\titlefoot{Niels Bohr Institute, Blegdamsvej 17,
     DK-2100 Copenhagen {\O}, Denmark
    \label{NBI}}
\titlefoot{IPNP of MFF, Charles Univ., Areal MFF,
     V Holesovickach 2, CZ-180 00, Praha 8, Czech Republic
    \label{NC}}
\titlefoot{NIKHEF, Postbus 41882, NL-1009 DB
     Amsterdam, The Netherlands
    \label{NIKHEF}}
\titlefoot{National Technical University, Physics Department,
     Zografou Campus, GR-15773 Athens, Greece
    \label{NTU-ATHENS}}
\titlefoot{Physics Department, University of Oslo, Blindern,
     NO-1000 Oslo 3, Norway
    \label{OSLO}}
\titlefoot{Dpto. Fisica, Univ. Oviedo, Avda. Calvo Sotelo
     s/n, ES-33007 Oviedo, Spain
    \label{OVIEDO}}
\titlefoot{Department of Physics, University of Oxford,
     Keble Road, Oxford OX1 3RH, UK
    \label{OXFORD}}
\titlefoot{Dipartimento di Fisica, Universit\`a di Padova and
     INFN, Via Marzolo 8, IT-35131 Padua, Italy
    \label{PADOVA}}
\titlefoot{Rutherford Appleton Laboratory, Chilton, Didcot
     OX11 OQX, UK
    \label{RAL}}
\titlefoot{Dipartimento di Fisica, Universit\`a di Roma II and
     INFN, Tor Vergata, IT-00173 Rome, Italy
    \label{ROMA2}}
\titlefoot{Dipartimento di Fisica, Universit\`a di Roma III and
     INFN, Via della Vasca Navale 84, IT-00146 Rome, Italy
    \label{ROMA3}}
\titlefoot{DAPNIA/Service de Physique des Particules,
     CEA-Saclay, FR-91191 Gif-sur-Yvette Cedex, France
    \label{SACLAY}}
\titlefoot{Instituto de Fisica de Cantabria (CSIC-UC), Avda.
     los Castros s/n, ES-39006 Santander, Spain
    \label{SANTANDER}}
\titlefoot{Dipartimento di Fisica, Universit\`a degli Studi di Roma
     La Sapienza, Piazzale Aldo Moro 2, IT-00185 Rome, Italy
    \label{SAPIENZA}}
\titlefoot{Inst. for High Energy Physics, Serpukov
     P.O. Box 35, Protvino, (Moscow Region), Russian Federation
    \label{SERPUKHOV}}
\titlefoot{J. Stefan Institute, Jamova 39, SI-1000 Ljubljana, Slovenia
     and Laboratory for Astroparticle Physics,\\
     \indent~~Nova Gorica Polytechnic, Kostanjeviska 16a, SI-5000 Nova Gorica, Slovenia, \\
     \indent~~and Department of Physics, University of Ljubljana,
     SI-1000 Ljubljana, Slovenia
    \label{SLOVENIJA}}
\titlefoot{Fysikum, Stockholm University,
     Box 6730, SE-113 85 Stockholm, Sweden
    \label{STOCKHOLM}}
\titlefoot{Dipartimento di Fisica Sperimentale, Universit\`a di
     Torino and INFN, Via P. Giuria 1, IT-10125 Turin, Italy
    \label{TORINO}}
\titlefoot{Dipartimento di Fisica, Universit\`a di Trieste and
     INFN, Via A. Valerio 2, IT-34127 Trieste, Italy \\
     \indent~~and Istituto di Fisica, Universit\`a di Udine,
     IT-33100 Udine, Italy
    \label{TU}}
\titlefoot{Univ. Federal do Rio de Janeiro, C.P. 68528
     Cidade Univ., Ilha do Fund\~ao
     BR-21945-970 Rio de Janeiro, Brazil
    \label{UFRJ}}
\titlefoot{Department of Radiation Sciences, University of
     Uppsala, P.O. Box 535, SE-751 21 Uppsala, Sweden
    \label{UPPSALA}}
\titlefoot{IFIC, Valencia-CSIC, and D.F.A.M.N., U. de Valencia,
     Avda. Dr. Moliner 50, ES-46100 Burjassot (Valencia), Spain
    \label{VALENCIA}}
\titlefoot{Institut f\"ur Hochenergiephysik, \"Osterr. Akad.
     d. Wissensch., Nikolsdorfergasse 18, AT-1050 Vienna, Austria
    \label{VIENNA}}
\titlefoot{Inst. Nuclear Studies and University of Warsaw, Ul.
     Hoza 69, PL-00681 Warsaw, Poland
    \label{WARSZAWA}}
\titlefoot{Fachbereich Physik, University of Wuppertal, Postfach
     100 127, DE-42097 Wuppertal, Germany
    \label{WUPPERTAL}}
\addtolength{\textheight}{-10mm}
\addtolength{\footskip}{5mm}
\clearpage
\headsep 30.0pt
\end{titlepage}
%
\pagenumbering{arabic} 
\setcounter{footnote}{0} %
\large
%

\newcommand{\dfrac}[2]{\frac{\displaystyle #1}{\displaystyle #2}}
\newcommand{\xx}      {\rule[-3mm]{0mm}{8mm}}
\def\qq{$q\overline{q}\;$}
\def\bb{$b\overline{b}\;$}
\def\cc{$c\overline{c}\;$}
\def\zqq{$Z \rightarrow q\overline{q}\;$}
\def\zbb{$Z \rightarrow b\overline{b}\;$}
\def\pis{$\pi_*\;$}
\def\pid{$\pi_{**}\;$}
\def\kad{$K_{**}$}
\def\dz{$D^0\;$}
\def\dzp{$D^{0/+}\;$}
\def\dplus{$D^+\;$}
\def\ds{$D^{*+}\;$}
\def\ddou{$D^{**}\;$}
\def\barb{$\bar{B}\;$}
\def\bdoul{$\bar{B} \rightarrow D^{**}\ell^-\bar{\nu}_{\ell}\;$}
\def\doudspi{$D^{**} \rightarrow D\pi_{**}\;$}
\def\doudspi2{$D^{**} \rightarrow D^{(*)}\pi_{**}\;$}
\def\ddpi{$D^{*+} \rightarrow D^0 \pi_*^+\;$}
\def\kpi{$K^-\pi^+\;$}
\def\k3pi{$K^-\pi^+\pi^+\pi^-$}
\def\kpipi0{$K^-\pi^+(\pi^0)\;$}
\def\dkpi{$D^0 \rightarrow K^-\pi^+\;$}
\def\dk3pi{$D^0 \rightarrow K^-\pi^+\pi^+\pi^-\;$}
\def\dkpipz{$D^0 \rightarrow K^-\pi^+(\pi^0)\;$}
\def\dkpipi{$D^+ \rightarrow K^-\pi^+\pi^+\;$}
\def\dskpi{$D^{*+} \rightarrow (K^-\pi^+)\pi_*^+\;$}
\def\dsk3pi{$D^{*+} \rightarrow (K^-\pi^+\pi^+\pi^-)\pi_*^+\;$}
\def\dskpipz{$D^{*+} \rightarrow (K^-\pi^+(\pi^0))\pi_*^+\;$}
\def\ddoupi{$D^{**} \rightarrow D \pi_{**}\;$}
\def\bcl{$b \rightarrow c \rightarrow \overline{\ell}\;$}
\def\bcbarl{$b \rightarrow \overline{c} \rightarrow \ell\;$}
\def\btaul{$b \rightarrow \tau \rightarrow \ell\;$}
\def\dm{$\Delta M\;$}
\def\mev{MeV/$c^2\;$}
\def\mevx{MeV/$c^2$}

\section{Introduction}
 
 The study of \barb meson semileptonic decays 
into any $D\pi$ or $D^{*}\pi$ final state is interesting for several reasons.
 Present measurements of \barb semileptonic decays into $D\ell^-\bar{\nu}_{\ell}$ and 
$D^*\ell^-\bar{\nu}_{\ell}$ imply that these final states account for 
only 60\% to 70\% of all \barb semileptonic decays~\cite{ref:PDG}.
The remaining contribution could be attributed to the production of 
higher excited states or non-resonant $D^{(*)}\pi$ final states,
hereafter denoted $D^{**}$.
However the ALEPH measurement of the $D^{**}\ell^-\bar{\nu}_{\ell}$ branching fraction 
does not fully account for the observed discrepancy~\cite{ref:ALEPH}.
 The ratio of branching fractions of \barb decays into $D^{*+}\pi^-\ell^-\bar{\nu}_{\ell}$
over all $D^{*+}\ell^-X$ final states~\footnote
{Throughout the paper charge-conjugate states are implicitly included;
 $\ell$ indicates an electron or a muon, not the sum over these two leptons.}
 is also a significant contribution to the systematic uncertainty 
on $\tau_{B^0_d}$, $\Delta m_d$ or $V_{cb}$ measurements~\cite{ref:PDG,ref:LEPHFS}.

 This paper describes a measurement of the branching fraction of  
\bdoul decays in the DELPHI experiment at LEP.
The decays of the $D^0$, \dplus and \ds into $D^0\pi_*^+$ are exclusively reconstructed~\footnote
{\pis denotes the charged pion from the $D^{*+} \rightarrow D^0\pi^+$ decay.}.
The analysis of \doudspi2 in \barb semileptonic decays~\footnote
{$D$ stands for $D^0$ or $D^+$;
\pid denotes the charged pion from the decay of a higher excited state of charmed meson
or from a non-resonant $D^{(*)}\pi$ final state.} 
relies on the impact parameter of the \pid candidate,
defined as its distance of closest approach to the reconstructed primary interaction vertex.
A similar technique has been applied previously in ALEPH~\cite{ref:ALEPHold,ref:ALEPH} and DELPHI~\cite{ref:VCB}.
The single pion final states $D^0\pi^+$, $D^+\pi^-$ or $D^{*+}\pi^-$, denoted ``right" sign, are expected to dominate
the decay widths. 
But pion pair emission, such as $D\pi^+\pi^-$,
is also allowed and could provide ``wrong" sign
$D^0\pi^-$, $D^+\pi^+$ or $D^{*+}\pi^+$ combinations.
Similarly $D_s$ orbitally excited states can decay into $D^0K^+$ or $D^{*0}K^+$ 
which can be distinguished from $D^{(*)0}\pi^+$ if the
kaon is identified.
 
The overall semileptonic branching fractions of a $b$ quark into $D^0$, \dplus or \ds final states
are also presented in this paper.

\section{The DELPHI detector}

The DELPHI detector has been described in detail elsewhere~\cite{ref:delphidet,ref:delphiperf}; 
only the detectors relevant to the present analysis are briefly described in the following.
The tracking of charged particles is accomplished in the barrel region with 
a set of cylindrical tracking detectors whose axis is oriented along the 
1.23~T magnetic field and the direction of the beam. 

 The Vertex Detector (VD) surrounds a Beryllium beam pipe with a radius of 5.5~cm.
It consists of three concentric layers
of silicon microstrip detectors at radii of 6, 9 and 11~cm from the beam line.
In 1991-1993 all the VD layers were single-sided with 
strips parallel to the beam direction. 
In 1994 and 1995, the innermost and the outermost layers were replaced by double-sided 
silicon microstrip modules,
providing a single hit precision of about 8~${\mu}$m in $r\phi$,
similar to that obtained previously, 
and between 10~${\mu}$m and 20~${\mu}$m in $z$~\cite{ref:vdpaper}~\footnote
{In the DELPHI coordinate system: $z$ is along the beam line, $\phi$ is the azimuthal angle in the $xy$ plane,
transverse to the beam axis, $r$ is the radius and $\theta$ is the polar angle with respect to the $z$ axis.}.
For polar angles between 44$^{\circ}$ and 136$^{\circ}$, a track crosses all the three VD layers.
The innermost layer covers the polar angle region between 25$^{\circ}$ and 155$^{\circ}$.
For charged particle tracks with hits in all three $r\phi$ VD layers,
the impact parameter precision is~\cite{ref:Chiara}: 
\begin{eqnarray}
 \sigma_{r\phi} = \dfrac{a}{p\sin^{3/2}\theta} \oplus b
\end{eqnarray}
where $a=61\pm1$~${\mu}$m, $b=20\pm1$~${\mu}$m and $p$ is the momentum in~GeV$/c$.

 The Inner Detector is placed outside the VD between radii of 12~cm and 28~cm. 
It consists of a jet chamber  giving up to 24 spatial measurements and a trigger chamber providing
a measurement of the $z$ coordinate.  The VD and ID are surrounded
by the main DELPHI tracking chamber, the Time Projection Chamber (TPC), which
provides up to 16 space points between radii of 30~cm and 122~cm. The Outer
Detector (OD) at a radius of 198~cm to 206~cm consists of five layers of drift cells. 
The  average momentum resolution of the tracking system is 
$\sigma(1/p)~<~1.5 \times 10^{-3}$~(GeV/$c)^{-1}$ 
in the polar angle region between 25$^{\circ}$ and 155$^{\circ}$.
The tracking in the forward ($11^{\circ}<\theta<33^{\circ}$)
and backward ($147^{\circ}<\theta<169^{\circ}$) regions 
is improved by two pairs of Forward drift Chambers (FCA and FCB) in the end-caps.

 Hadrons were identified using the specific ionization ($dE/dx$) in the TPC
and the Cherenkov radiation in the barrel Ring
Imaging CHerenkov detector (RICH) placed between the TPC and the OD detectors.

 The muon identification  relied mainly on the muon chambers, 
a set of drift chambers giving three-dimensional information
situated at the periphery of DELPHI after approximately 1~m of iron.

 Electron identification relied mainly on the electromagnetic calorimeter in the barrel region 
(High density Projection Chamber HPC) which is a sampling device having a relative energy 
resolution of $\pm$5.5\% for electrons with 45.6~GeV$/c$ momentum, and a spatial resolution 
along the beam axis of $\pm$2~mm.

\section{Event selection and simulation}
 
 Charged particles were required to have 
a measured momentum between 0.3~GeV$/c$ and 50~GeV$/c$, a relative error
on momentum less than 100\%, a  track length in the 
TPC larger than 30~cm and a distance of closest approach to
the interaction point of less than 4~cm in $r$ and less than 10~cm in $z$.

 Hadronic events were required to have at least five charged particles 
with momentum greater than 0.4~GeV$/c$ and a
total energy of the charged particles (assumed to be pions) greater than 12\%
of the collision energy.
 A total of $N_Z =3.51$ million hadronic events was obtained from the 1992-1995 data.
 Simulated hadronic events were generated using the JETSET~7.3 Parton Shower program~\cite{ref:lund}:
8.5 million \zqq and 4.0 million \zbb generated events,
corresponding to seven times the available statistics in real data for \bb final states.
The $B$ meson mean lifetime was set to $\tau_B^{\mathrm MC}=1.6$~ps.
The generated events were followed through a detailed detector simulation~\cite{ref:delphiperf}
and then processed through the same analysis chain as the real data. 
The hadronic event selection efficiency was thus estimated to be $\epsilon_Z=95.7\%$.
The data sample contained also $0.2\%$ of $\tau$ pair events and $0.2\%$ of Bhabha events.

 The primary interaction vertex was computed in space for each event using an iterative procedure
based on the $\chi^2$ of the fit.
The average transverse position of the interaction point, known for each fill, 
was included as a constraint during the primary vertex fit. 
In order to increase the \bb purity of the selected sample,
using the impact parameter of all measured charged particle tracks in the event,
the probability that all these tracks originate from the primary vertex
was required to be smaller than 0.1~\cite{ref:Btag}.
This selection retains $15\%$ of $Z \rightarrow u\bar{u}$, $d\bar{d}$ and $s\bar{s}$ events,
$48\%$ of $Z \rightarrow c\bar{c}$ events and $94\%$ of $Z \rightarrow b\bar{b}$ events.

 In order to estimate the reconstruction efficiencies and the invariant mass resolutions,
dedicated samples of events containing a \barb meson decaying into 
$D^0\pi^+\ell^-\bar{\nu}_{\ell}X$, $D^+\pi^-\ell^-\bar{\nu}_{\ell}X$ or $D^{*+}\pi^-\ell^-\bar{\nu}_{\ell}X$ were generated. 
Physical backgrounds have also to be studied. These can be due to $b \rightarrow c W^-$ decays followed
by $W^- \rightarrow \overline{c} s$ and $\overline{c} \rightarrow \ell^-\bar{\nu}_{\ell}X$
(hereafter denoted \bcbarl background),
or followed by $c \rightarrow \ell^+\nu_{\ell}X$ with $W^- \rightarrow \bar{D}X$ (denoted \bcl background).
For this purpose,
some dedicated samples of $\bar{B} \rightarrow D \bar{D}_s X$ or $D \bar{D} K X$ decays,
with the $\bar{D}_s$ or $\bar{D}$ decaying semileptonically, were generated.

\section{$D^{(*)}\ell^-$ selection}
 
\subsection{Lepton selection and identification}

 Both muon and electron candidates were selected with a momentum larger than 2~GeV$/c$.
The lepton candidate was required to have at least one hit associated in the Vertex Detector.

The muon identification algorithm is described in reference~\cite{ref:delphiperf}.
A ``loose" selection criterion provided an identification
efficiency of $(90\pm2)\%$ for a probability of misidentifying a charged hadron as a muon of 1.2\% 
within the acceptance of the muon chambers.

A neural network procedure, combining information from several detectors,
has been developed for electron identification.
Electrons were identified with an efficiency of $(65\pm2)\%$ and a
misidentification probability that a hadron be identified as electron of about 0.4\%~\cite{ref:pcdstar}.

 The lepton transverse momentum relative to the $D^{(*)}$ meson momentum vector (as defined below) 
was required to be larger than 0.7~GeV$/c$.
This cut reduced the contamination of leptons from $b$ semileptonic decay into $\tau$ 
and from \bcbarl or \bcl decays.

\subsection{$D^{(*)}$ decay channels}

 The $D^{(*)}$ meson candidates were reconstructed in the following decay channels: 
\dkpi or \k3pi (for $D ^0$ not coming from a \ds decay), 
\dkpipi and \ddpi with a \dz decaying into
$K^-\pi^+$, \k3pi or \kpipi0 where the $\pi^0$ was not reconstructed. 
In order to optimize the statistical precision of the measured production rates,
slightly different selection criteria, as described below, were chosen in each $D^{(*)}$ meson sample.

 Only charged particles with momentum vectors in the hemisphere defined by the lepton direction were considered
for the reconstruction of charmed mesons.
The kaon candidate from the $D$ decay was required to have the same charge as the identified lepton.
The kaon momentum was required to be larger than 1 (2)~GeV$/c$ in the \dz ($D^+$) channel.
The momentum of each pion from the \dzp decay had to be larger than 1~GeV$/c$, except
for the \k3pi final state where the minimum momentum
of candidate pions was lowered to 0.5 (0.3)~GeV$/c$ in the \dz ($D^{*+}$) analysis. 
Any charged particle
with a momentum between 0.3~GeV$/c$ and 4.5~GeV$/c$ and a charge opposite to that of the kaon
was used as pion candidate for the \ddpi decay channel. 

 To reduce the combinatorial background for all channels, 
except in the $D^{*+} \rightarrow (K^-\pi^+)\pi_*^+$ decay,
the kaon candidate of the $D$  was required to be identified according to the RICH and $dE/dx$ information~\cite{ref:delphiperf}.
In the \dkpi and \dkpipz decay channels, the angle $\theta^*$ between 
the $K^-\pi^+$ momentum vector and the kaon direction in the $K^-\pi^+$ rest frame was required to satisfy the condition
$\cos\theta^*>-0.8$.
For genuine \dz candidates an isotropic distribution in $\cos\theta^*$ 
is expected whereas the background is strongly peaked
in the backward direction. 

 The $\pi_{*}$ candidate and at least
two particles from the \dzp decay were required to have at least one hit associated in the Vertex Detector.

\subsection{$D$ vertex}

 After the previous selections, a $K^-\pi^+$, $K^-\pi^+\pi^+$ or \k3pi vertex was fitted in space.
In the \k3pi decay channel, 
in order to reduce the large combinatorial background,
the impact parameters of charged particle trajectories,
relative to the common \dz vertex, were required
to be smaller than 150~$\mu{\mathrm m}$.
In the \dkpipi channel, either these impact parameters had to be less than 100~$\mu{\mathrm m}$
or the $\chi^2$ probability of the $K^-\pi^+\pi^+$ vertex had to be larger than $10^{-4}$. 

 The momentum vector of each particle, attached to the $D$ vertex, was recomputed at this vertex.
In each channel, the scaled $D$ energy,
$X_E(D) = E(D) / E_{\mathrm beam}$, was required to be larger than 0.15.
 
 The apparent decay length of the \dz or \dplus candidate, $\Delta L$, 
was computed in the plane transverse to the beam axis.
It was given the same sign as the scalar product of the $D$ momentum direction
with the vector joining the primary to the $D$ vertices.
In the \ds channel, $\Delta L$ was required to be positive.
In the \dz and \dplus channels, which have a higher combinatorial background,
the value of $\Delta L$ divided by its error was required to be larger than 1.

\subsection{$B$ vertex}

 Finally a $D^0\ell$, $D^+\ell$ or $D^0\pi_*\ell$ vertex (denoted ``$B$" vertex in the following) was fitted in space.
The $B$ decay length
was defined, as above, as the signed distance between the primary vertex and the secondary $D(\pi_*)\ell$ vertex
in the plane transverse to the beam axis.
This $B$ decay length divided by its error was required to be larger than 1 for all channels.
In order to reduce further the combinatorial background,
the decay length divided by its error between the $D$ and the $B$ vertices was also computed:
it was required to be larger than -1 in the \dz samples 
and in the $D^{*+} \rightarrow (K^-\pi^+\pi^+\pi^-)\pi_*^+$ sample,
and to be positive in the \dplus sample.

\subsection{$D$ invariant mass}

 The selection of $D^{*+}\ell^-$ events relied on the small mass difference 
($\Delta M$) between the \ds and the candidate $D^0$.
On the contrary, \ds candidates were rejected from the \dz and \dplus samples as follows:
the \dz candidates were rejected if at least one \pis particle
was found in the event giving a \dm value less than 160~\mevx;
in the \dkpipi sample, both \kpi pairs were associated to the remaining $\pi^+$
and a \dm mass difference was computed, 
the $K^-\pi^+\pi^+$ combination was rejected if at least one \dm value was found smaller than 160~\mevx.

 Figure~\ref{fig:dmass} shows the invariant mass (or mass difference) distributions 
in each of the previously selected $D$ meson channels.
In the \dskpi channel, the \kpi invariant mass had to be within 75~\mev of the nominal \dz mass.
In the \dskpipz channel, the \kpi invariant mass had to be between 1500 and 1700~\mevx.
In the \dsk3pi channel, the \dm mass difference had to be within 2~MeV$/c^2$ of 
the nominal $D^{*+}-D^0$ mass difference. 
The invariant mass of the \dkpi channel
has a resolution of about 25~\mev whereas it is less than 15~\mev in the \k3pi final state.
Thus, in the particular case of the \dskpi decay channel, 
the \kpi invariant mass was constrained to the \dz mass value
and a constrained ($D^0\pi_*\ell$) kinematic fit was performed.
This improved the resolution of the ($D^{*+}-D^0$) mass difference by 30\% in this channel.
 
 A clear signal corresponding to $D\ell^-$ events is observed in each distribution
(data points), whereas the wrong sign $D\ell^+$ combinations (hatched histograms)
present a much smaller $D$ meson contribution.
The right sign invariant mass distributions were fitted with a signal component 
described by the sum of Gaussian functions, and a combinatorial background parameterised
with a polynomial form.
In the $D^0 \rightarrow K\pi$ and $K\pi\pi\pi$ samples, the contribution from missing $\pi^0$ 
appears as a ``satellite" peak for mass values smaller than the nominal \dz mass. 
This contribution was parameterised as the sum of Gaussian functions
with their parameters fixed according to the simulation.
In each channel,
the relative amounts and relative widths of the Gaussian functions 
describing the $D$ signal were tuned according to the simulation.
The free parameters of the fits were thus
the coefficients of the polynomial background, 
the normalisation of the ``satellite" peak 
(in the $K\pi$ and $K\pi\pi\pi$ invariant mass distributions), 
the average width and mean value of the signal shape 
and the number of $D$ signal candidates.
For each decay channel, the mass distributions of the wrong sign $D\ell^+$ events were
fitted with the same shape parameters as the right sign signals.
This allowed the contribution of the fake lepton events to be determined and then subtracted. 
The observed numbers of $D$ mesons,
within the quoted range around the \dzp mass and $D^{*+}-D^0$ mass difference, is indicated in Table~\ref{tab:dyield}.
\begin{table}[bth]
\begin{center}
\begin{tabular}{|c||c|c||c|c|} \hline
 \xx                  & \multicolumn{2}{c|}{Mass range (MeV$/c^2$)} & Nb. of             & Nb. of    \\ 
 $D$ sample           & $M(D^{0/+}$)      & \dm                     & $D\ell^-$          & $D\ell^+$ \\ \hline\hline
 \dkpi                & 1820-1910         & $>$160                  & 752$\pm$41         & ~6$\pm$18 \\ 
 \dk3pi               & 1840-1890         & $>$160                  & 689$\pm$43         & 39$\pm$26 \\ \hline
 \dkpipi              & 1830-1910         & $>$160                  & 763$\pm$44         & 66$\pm$19 \\ \hline
 \dskpi               & 1790-1940         & 143.5-147.5             & 416$\pm$24         & 18$\pm$5  \\ 
 \dsk3pi              & 1840-1890         & 143.5-147.5             & 303$\pm$21         &  5$\pm$5  \\ 
 \dskpipz             & 1500-1700         & $<$155                  & 522$\pm$33         & 15$\pm$12 \\
\hline 
\end{tabular}
\caption[]{Mass selections and number of $D$ candidates observed in each decay channel (with their statistical error). 
Note that most of the \ddpi candidates were removed from the selected \dz sample;
the \dplus sample also includes $D^{*+}\rightarrow D^+ \pi^0$ (or $\gamma$) decays.
}
\label{tab:dyield}
\end{center}
\end{table}

\section{Semileptonic $b$ decay rate into $D \pi \ell^- X$ or $D^* \pi \ell^- X$}

 In this section, a search for any $D\pi_{**}$ final state is described,
based on the impact parameter distribution of the \pid candidates relative to the primary interaction vertex
and using the $D$ decay channels selected in the previous section.

\subsection{$D^{**}\ell^-$ selection}
 
 The selection criteria for the additional \pid candidate were identical for all decay channels.
All charged particles with a momentum greater than 0.5~GeV$/c$ 
and produced in the hemisphere defined by the $D(\pi_*)\ell^-$ momentum vector 
were considered as \pid candidates.
The invariant $D(\pi_*)\pi_{**}\ell$ mass had to be smaller than 5.5~GeV$/c^2$.
The \pid track was required to have at least 2 hits in the Vertex Detector.
Its combined RICH and $dE/dx$ information had not to be compatible with the kaon hypothesis.
The impact parameter of this \pid relative to the previously fitted $D(\pi_*)\ell$ vertex
was required to be smaller than 100~$\mu$m.

 For each \pid candidate,
the impact parameter relative to the primary interaction vertex was computed in the plane transverse to the beam axis.
The sign of this impact parameter was defined with respect to the $D(\pi_*)\ell$ direction.
It was positive if the intercept between the \pid and the $D(\pi_*)\ell$ momentum vectors was
downstream of the primary vertex along the $D(\pi_*)\ell$ direction, and negative if it was upstream~\cite{ref:Btag}.

\begin{figure}[bth]
\epsfverbosetrue
\begin{center}
\mbox{\epsfxsize=17cm,
\epsffile{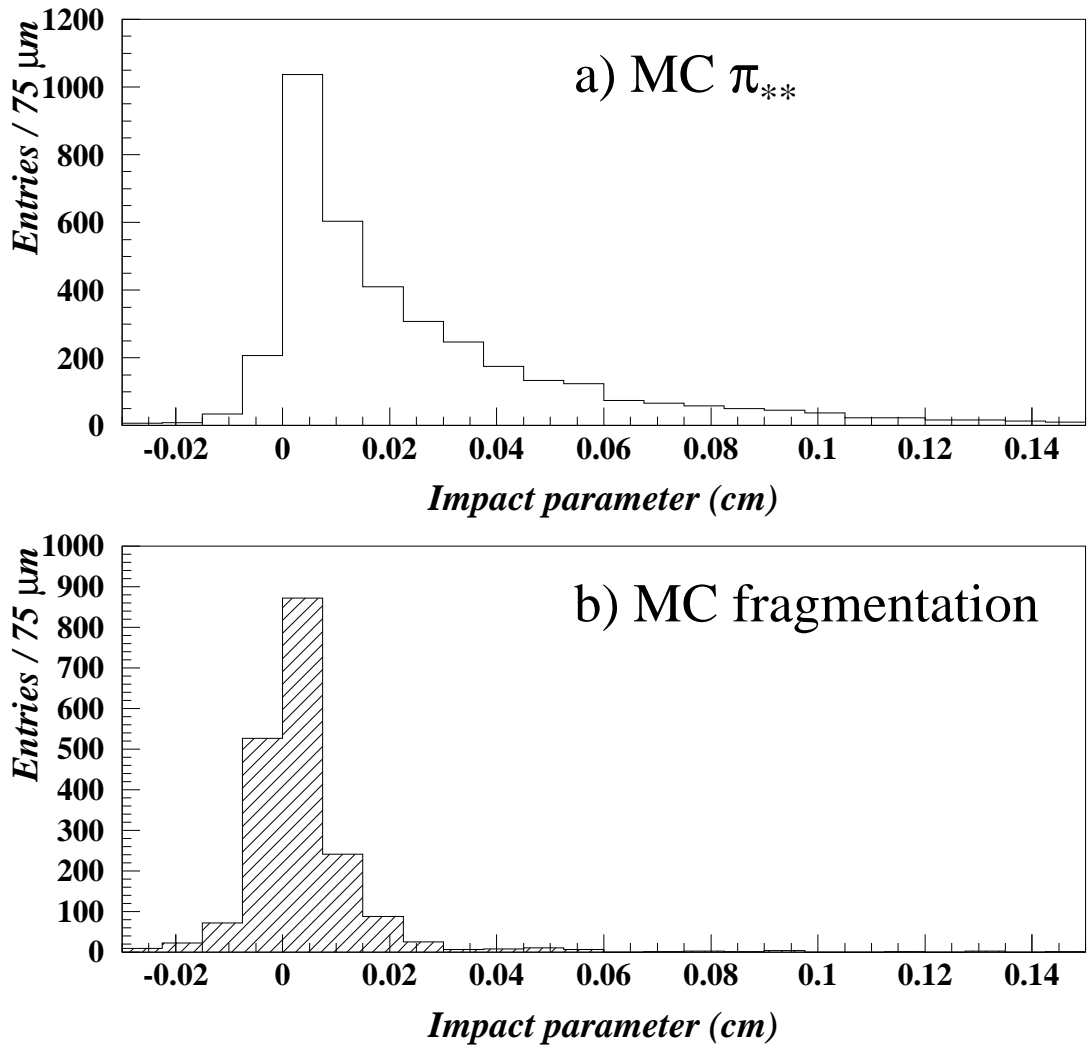}}
\vspace*{-1.0cm}
\caption[] {
Impact parameter relative to the primary interaction vertex in simulated $B$ semileptonic decays for 
a) \pid from \ddou decay 
(using a $B$ mean lifetime value of 1.6~ps)
and b) charged particles from jet fragmentation (see Section~5.1).}
\label{fig:mcimpa} 
\end{center}
\end{figure}

 The \pid impact parameter distribution of simulated $B$ semileptonic decays is shown in Figure~\ref{fig:mcimpa}a.
Compared with charged particles 
produced in $b$ quark fragmentation or gluon radiation in jets (see Figure~\ref{fig:mcimpa}b),
\pid from \bdoul decays
present a tail at large positive impact parameters due to the long $B$ lifetime.

\subsection{Backgrounds}
 
\begin{figure}[bth]
\epsfverbosetrue
\begin{center}
\mbox{\epsfxsize=17cm,
\epsffile{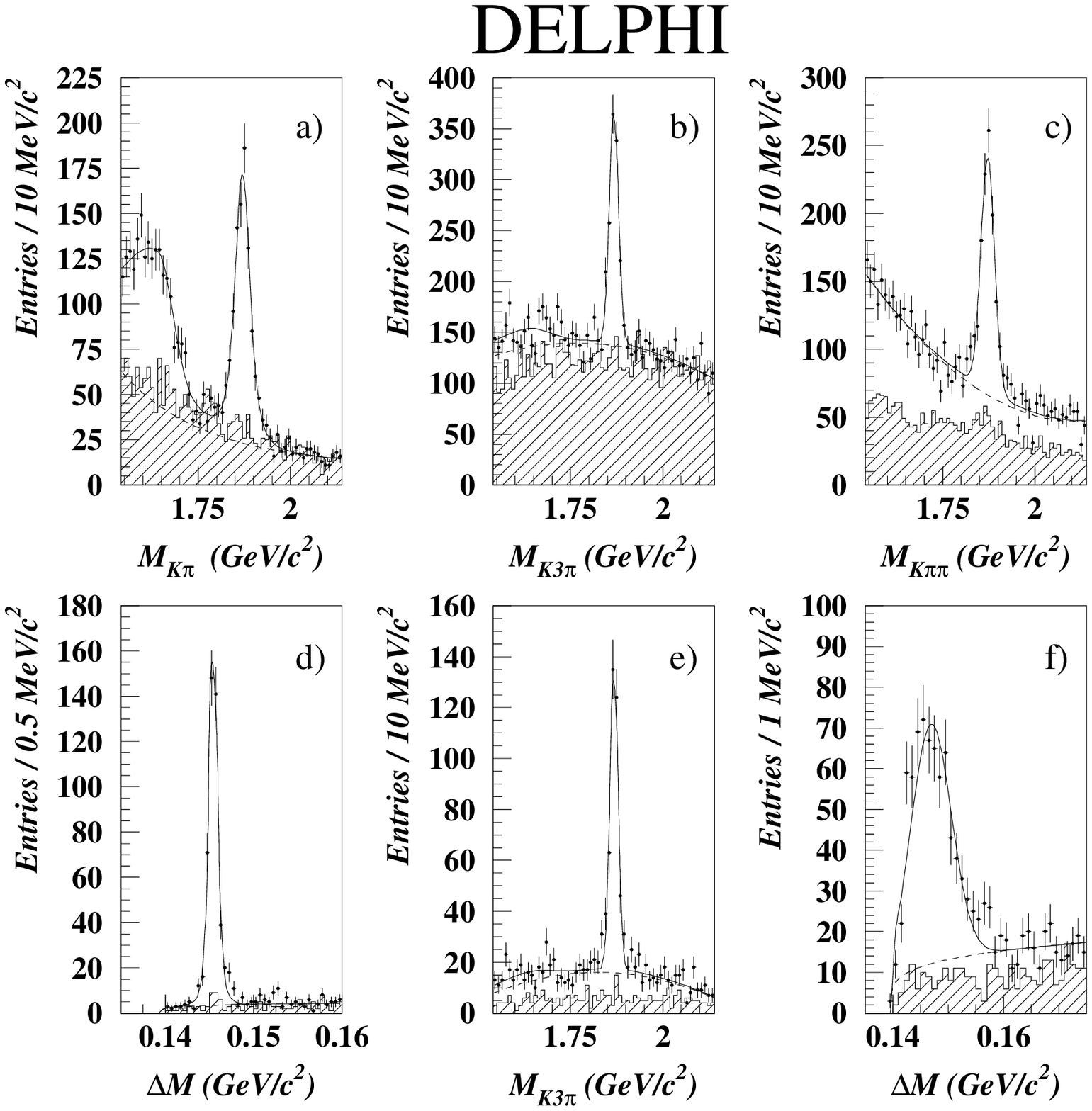}}
\vspace*{-1.0cm}
\caption[] {Invariant mass distributions in the
(a) \dkpi (b) \dk3pi (c) \dkpipi and (e) \dsk3pi decay channels;
mass difference distributions $M(K^-\pi^+\pi_*^+)-M(K^-\pi^+)$ in the
(d) \dskpi and (f) \dskpipz decay channels.
The reconstructed \ds candidates have been removed in a,b,c.
Right charge $D\ell^-$ (dots) and wrong charge $D\ell^+$ (hatched histogram) events are shown.
The solid line curves are fits which include a background parameterisation (dashed curve alone)
and Gaussian functions for the signal (see Section~4).
}
\label{fig:dmass} 
\end{center}
\end{figure}

\begin{figure}[bth]
\epsfverbosetrue
\begin{center}
\mbox{\epsfxsize=17cm,
\epsffile{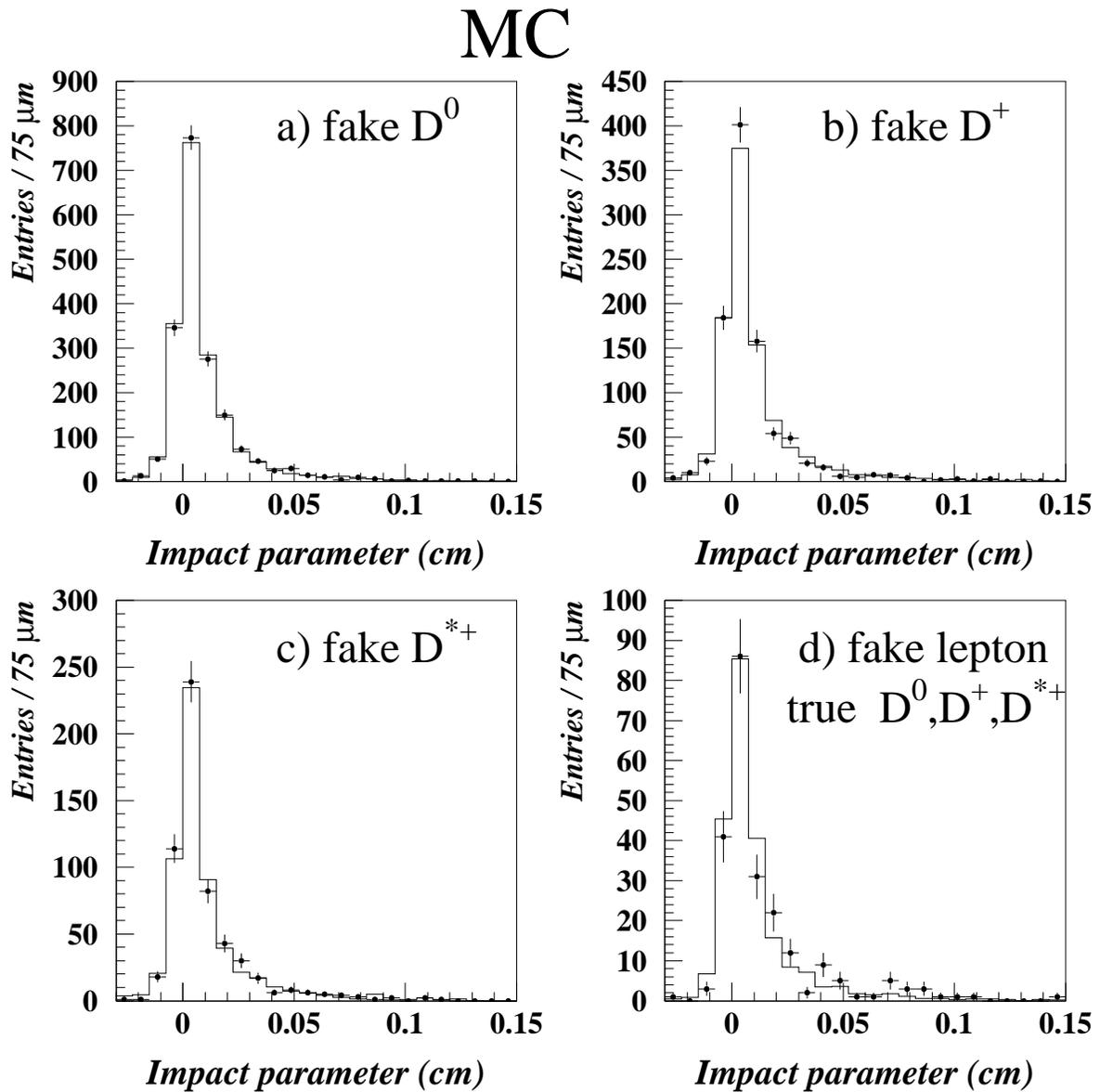}}
\vspace*{-1.0cm}
\caption[] {In the simulation: impact parameter distributions of pions accompanying (a-c) a fake $D$ meson 
(points with error bars)
or a $D$ selected in the tails of the mass distributions (histograms);
(d) a fake lepton (points with error bars)
or a $D\ell^+$ where the $D$ was selected in the signal region (histograms, see Section~5.2).}
\label{fig:fakedl} 
\end{center}
\end{figure}

 For real data $D^{**}\ell$ candidates, two sources of background had to be subtracted:
\begin{enumerate}
\item [$\bullet$] Fake $D$ associated to a lepton candidate:
this combinatorial background was estimated by using events in the tails of the mass distributions of Figure~\ref{fig:dmass},
after a normalisation to the fraction of events below the $D$ signals.
Figures~\ref{fig:fakedl}a-c present the impact parameter distributions of all pion candidates
associated to a fake $D$ (points with error bars) and to a $D$ in the tails of the mass distributions (histograms)
from the $D$ samples selected in the \qq simulation.
A good agreement is found between the true background and the mass tail estimate.
\item [$\bullet$] True $D$ associated to a fake lepton:
this background is due to charged pions and kaons misidentified as leptons. 
It has been subtracted by using the $\pi_{**}$ candidates produced in the same direction as a wrong sign $D\ell^+$ event
(shown in the hatched histograms of Figure~\ref{fig:dmass})
where the $D$ candidate was selected in the mass range defined in Table~\ref{tab:dyield}.
Figure~\ref{fig:fakedl}d presents the impact parameter distributions of all pion candidates
associated to a fake lepton (points with error bars) and to a $D\ell^+$ event (histograms)
from the $D$ samples selected in the \qq simulation.
Here also a good agreement is found between the true background and the $\ell^+$ estimate.
\end{enumerate}
In the real data events, the same procedure was applied.
The shapes of these backgrounds were taken from the real data themselves and their 
normalisation was estimated according to the 
fit of the mass distributions of Figure~\ref{fig:dmass}.

After the subtraction of these backgrounds,
all the remaining pions can be attributed to $b$ decays into $D\pi\ell^-X$ final state.
However, four kinds of pions are still to be considered:
\begin{enumerate}
\item [$\bullet$] genuine \pid from \bdoul decays (see Figure~\ref{fig:mcimpa}a);
\item [$\bullet$] particles from jet fragmentation (see Figure~\ref{fig:mcimpa}b);
\item [$\bullet$] ``$\tau,c\rightarrow\ell$" background: 
it includes pions produced in \ddou decays when the \ddou is not issued
from a direct semileptonic ($e$ or $\mu$) $b$ decay.
This \ddou can be produced in $b \rightarrow D^{**}\tau^-\bar{\nu}_{\tau}$ decay,
or in $b \rightarrow D^{**} \bar{D}_{(s)} X$ (or $b \rightarrow D \bar{D}^{**} X$) transitions,
when the other $\bar{D}_{(s)}$ (or $D$) meson decays semileptonically;
\item [$\bullet$] ``hadronic" background: 
it is due to other hadrons, denoted $H$, 
produced from the $\overline{c}$ in \bcbarl decay events or from the $c$ in \bcl 
(when the other charm quark fragments into a $D$ meson).
Such hadrons can be also emitted directly from the $\bar{B}$ meson.
\end{enumerate}
Despite the momentum and transverse momentum cuts applied to the lepton, these last two classes were not fully eliminated.
Their impact parameter distributions were
similar to the impact parameter distribution of genuine \pid from $b$ semileptonic decays.
These two last backgrounds were thus fitted together with the genuine \pid signal and subtracted only afterwards.
Measured results were used for their rates
and their selection efficiencies were obtained from the simulation (see Section 5.4.3).

\subsection{Total yield}

\begin{figure}[bth]
\epsfverbosetrue
\begin{center}
\mbox{\epsfxsize=17cm,
\epsffile{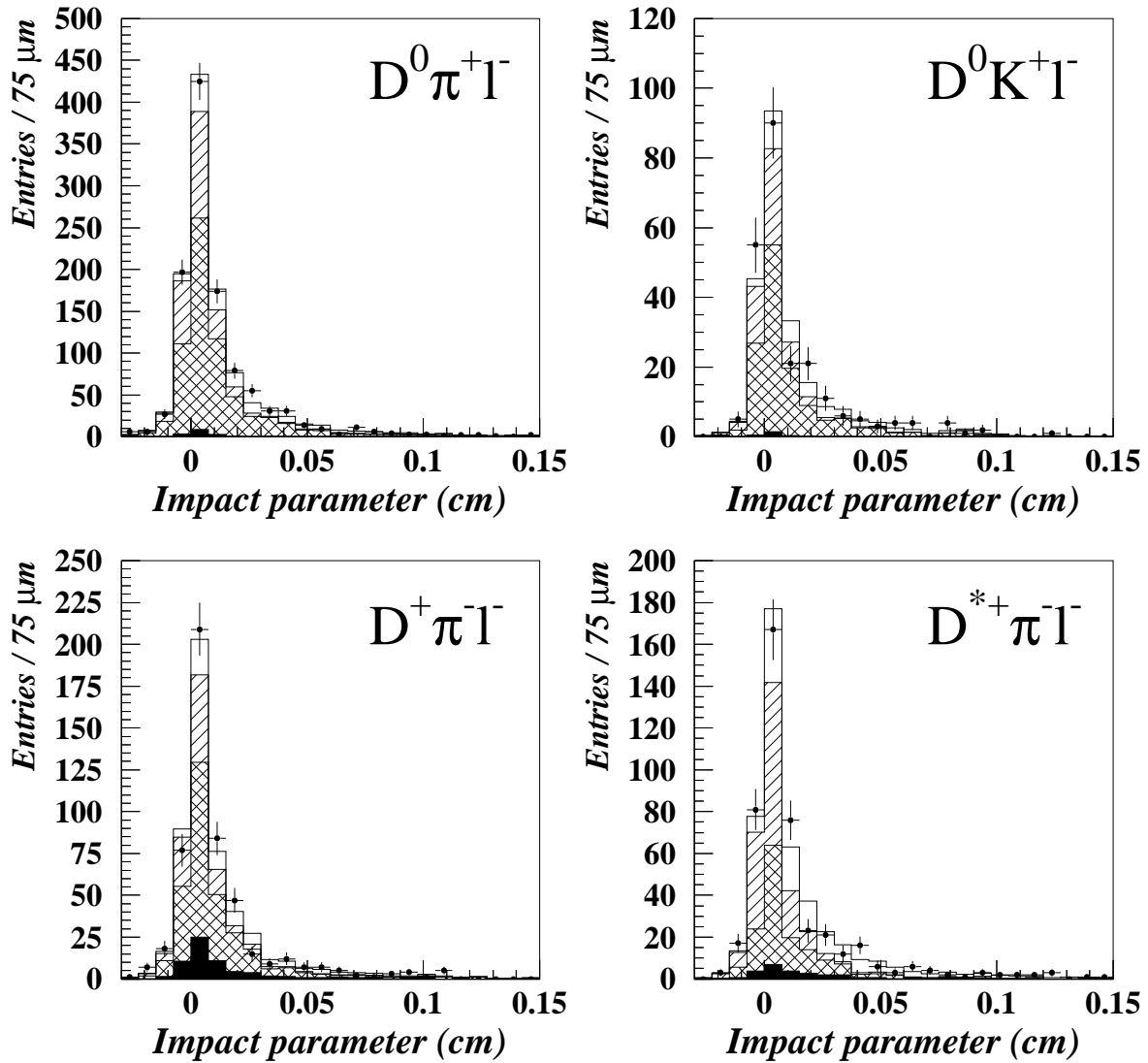}}
\vspace*{-1.0cm}
\caption[] {
Impact parameter relative to the primary interaction vertex in real data for 
``right" sign $D^0\pi^+$, $D^0K^+$, $D^+\pi^-$ and $D^{*+}\pi^-$ candidates.
The black and cross-hatched histograms are the estimated contributions from fake leptons
and fake $D$ mesons, respectively.
The hatched and empty area histograms are the fitted contributions from jet fragmentation
and \pid from \ddou decays, respectively (see Section~5.3).}
\label{fig:imptot}
\end{center} 
\end{figure}

\begin{figure}[bth]
\epsfverbosetrue
\begin{center}
\mbox{\epsfxsize=17cm,
\epsffile{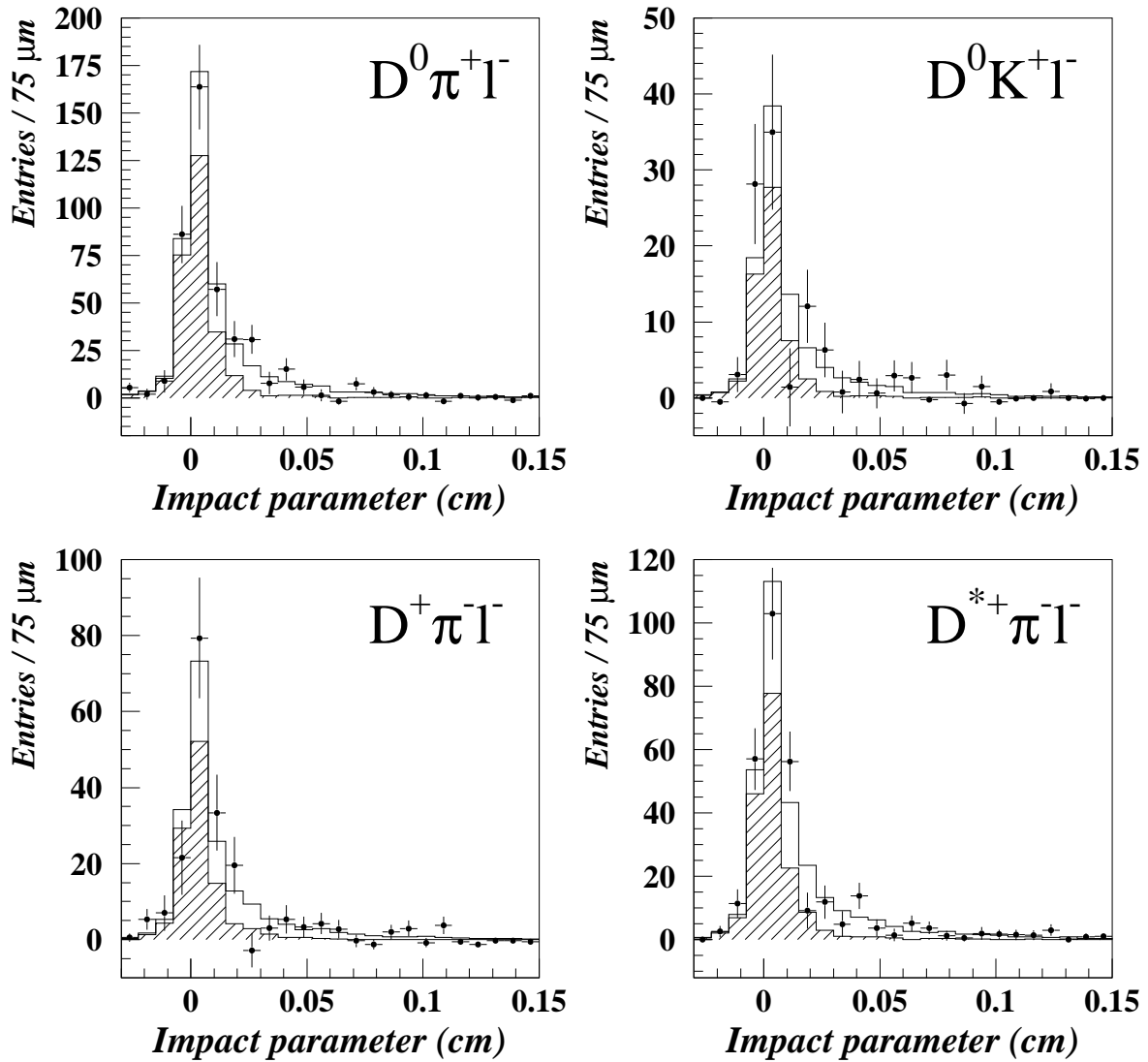}}
\vspace*{-1.0cm}
\caption[] {
Impact parameter relative to the primary interaction vertex in real data for 
background subtracted ``right" sign $D^0\pi^+$, $D^0K^+$, $D^+\pi^-$ and $D^{*+}\pi^-$ candidates.
The hatched and empty area histograms are the fitted contributions from jet fragmentation
and \pid from \ddou decays, respectively (see Section~5.3).}
\label{fig:impars}
\end{center} 
\end{figure}

\begin{figure}[bth]
\epsfverbosetrue
\begin{center}
\mbox{\epsfxsize=17cm,
\epsffile{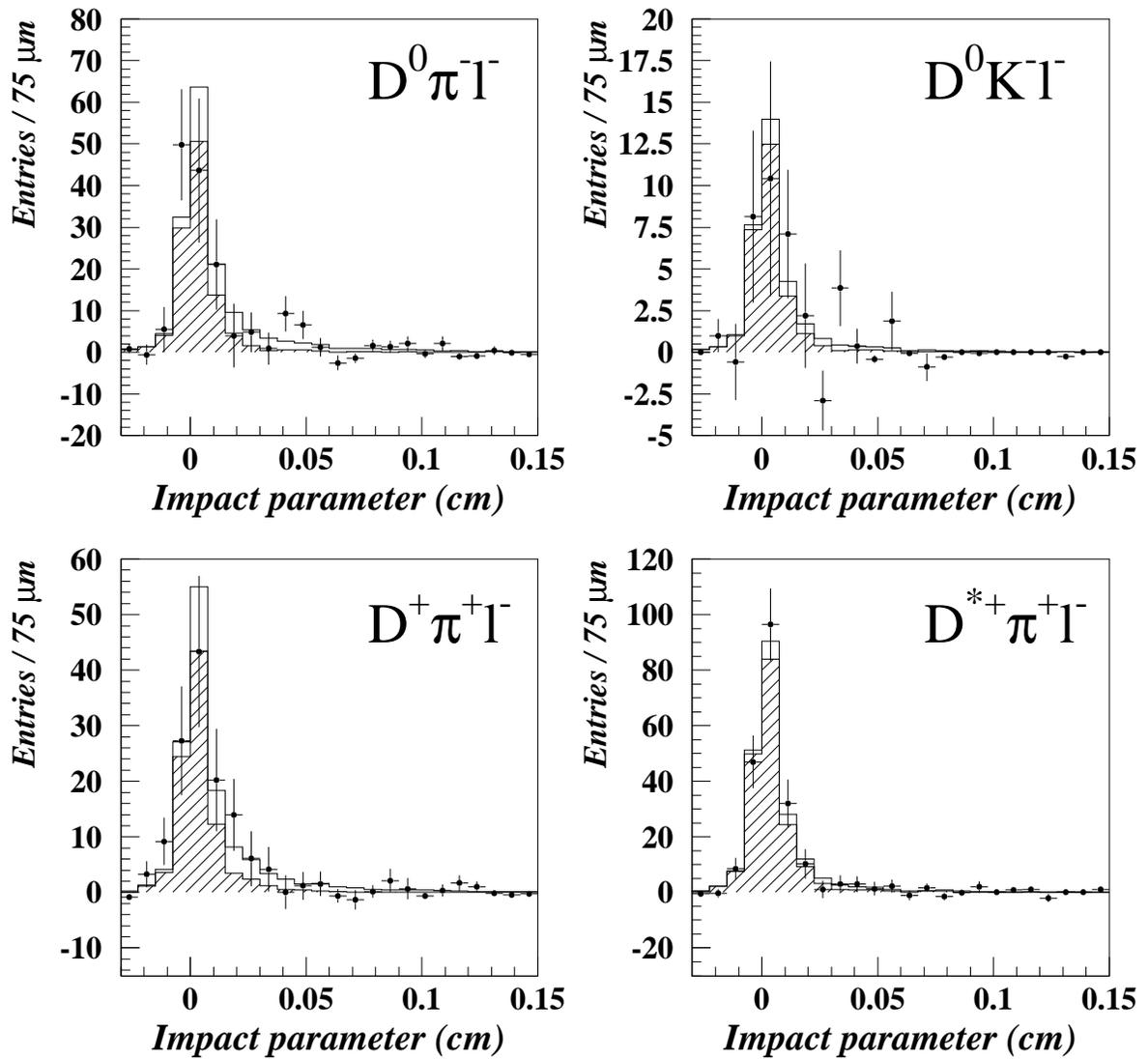}}
\vspace*{-1.0cm}
\caption[] {
Same as Figure~\ref{fig:impars} for
background subtracted ``wrong" sign $D^0\pi^-$, $D^0K^-$, $D^+\pi^+$ and $D^{*+}\pi^+$ candidates.}
\label{fig:impaws}
\end{center} 
\end{figure}

 In the real data, 
the impact parameter distributions of the \pid candidates of the
``right" sign $D^0\pi^+\ell^-$, $D^+\pi^-\ell^-$ and $D^{*+}\pi^-\ell^-$  samples
are shown in Figure~\ref{fig:imptot}.
They were fitted, fixing the fake $D$ and fake lepton backgrounds, but letting free the normalisation of the 
fragmentation and \pid components.
Figure~\ref{fig:impars} shows the same distributions, after subtraction of the fake $D$ and fake lepton backgrounds.
Similar fits were performed to the ``wrong" sign $D^0\pi^-\ell^-$, $D^+\pi^+\ell^-$ and $D^{*+}\pi^+\ell^-$ 
samples and are shown in Figure~\ref{fig:impaws} after the subtraction of the fake $D$ and fake lepton backgrounds.

 Instead of rejecting kaons in order to select \pid, 
kaons were also identified in order to select $K_{**}$
from $D_{s1} \rightarrow D^{*0} K^+$, $D_{s2}^* \rightarrow D^0 K^+$ decays or any other $D^0 K^+ X$ final state
from other $D_{sJ}$ resonances. 
The corresponding impact parameter distributions are shown in the same figures as above.
The kaon rejection requirement led to a \pid identification efficiency of $(92\pm1)\%$ and a probability
of wrong assignment as a kaon of $(8\pm1)\%$.
The kaon identification requirement lead to a $K_{**}$ identification efficiency of $(60\pm2)\%$ and a probability
of wrong assignment as a pion of $(40\pm2)\%$.
These factors were obtained from the real data, as explained in Section~5.4.1 ($f_{\pi{\mathrm id}}$ correction).
The numbers of fitted $D^0\pi\ell^-$ and $D^0K\ell^-$ were $163\pm34(stat)$ and $39\pm15(stat)$ 
($48\pm21(stat)$ and $5\pm8(stat)$) in the ``right" sign (``wrong" sign) samples. 
These values need to be corrected in order to take into account the fraction of kaons misidentified as pions.

 The final amounts, $N(D\pi\ell^-)$, of ``right" and ``wrong" signs fitted candidates are presented in Table~\ref{tab:piyield} 
for all considered channels.
In the \dz channel, the separated contributions of $D^0\pi$, $D^0K$ and the total $D^0h$ are also indicated
(where ``$h$" means that the \pid candidate was selected without identification).
Significant numbers of ``right" sign candidates are fitted for all channels, 
whereas the number of ``wrong" signs are clearly smaller.

\begin{table}[bth]
\begin{center}
\begin{tabular}{|l||c|c|c|c|c|} \hline
 Sample                   &  $D^0h\ell^-$  & $D^0\pi\ell^-$ &  $D^0K\ell^-$ &  $D^+\pi\ell^-$ &  $D^{*+}\pi\ell^-$  
\\ \hline\hline
 $N$(``right" sign) 
                          &  $202\pm37$     &  $182\pm39$   &  $20\pm18$    &  $75\pm25$      &  $132\pm22$   \\
 $N$(``wrong" sign) 
                          &  ~$53\pm23$     &  ~$55\pm24$   &  $-2\pm10$    &  $41\pm20$      &  ~$24\pm16$   \\ \hline\hline
 $\epsilon_{D\ell}$     
                          & \multicolumn{3}{c|} {$0.127\pm0.002$}  &  $0.095\pm0.002$  &  $0.150\pm0.002$  \\
 $\epsilon_{**}$   
                          & \multicolumn{3}{c|}{$0.655\pm0.006$}   & $0.649\pm0.008$   & $0.654\pm0.005$ \\ \hline
 $f_{K\pi}$ 
                          & \multicolumn{3}{c|} {$1.87\pm0.09$}    &  $1$              &  $3.02\pm0.16$    \\ 
 $f_{M_D}$ 
                          & \multicolumn{3}{c|} {$0.94\pm0.01$}    &  $0.98\pm0.01$    &  $1.01\pm0.01$    \\ 
 $f_{\mathrm VD}$      
                          & \multicolumn{3}{c|} {$1.00\pm0.03$}    &  $1.00\pm0.03$    &  $1.00\pm0.03$    \\ 
 $f_{D{\mathrm vtx}}$      
                          & \multicolumn{3}{c|} {$1$}              &  $0.97\pm0.03$    &  $1$    \\ 
 $f_{K{\mathrm id}}$ 
                          & \multicolumn{3}{c|} {$0.84\pm0.02$}    &  $0.83\pm0.02$    &  $1$    \\ 
 $f_{\pi{\mathrm id}}$      
                          & \multicolumn{3}{c|} {$1$}              &  $0.92\pm0.01$    &  $0.92\pm0.01$  \\ \hline
 $f_{\tau_B}$(``right" sign) 
                          & \multicolumn{3}{c|} {$1.02\pm0.02$}    &  $0.98\pm0.02$    &  $0.98\pm0.01$    \\ 
 $f_{\tau_B}$(``wrong" sign) 
                          & \multicolumn{3}{c|} {$0.98\pm0.02$}    &  $1.02\pm0.02$    &  $1.03\pm0.03$    \\ \hline 
 $f_{\tau,c\rightarrow\ell}$   
                          & \multicolumn{3}{c|} {$0.075\pm0.030$}  &  $0.075\pm0.030$  &  $0.075\pm0.030$    \\ \hline
 ${\cal F}_{H}$ ($\times 10^{-3}$)   
                          & $1.06\pm0.29$ & $0.78\pm0.22$ & $0.28\pm0.09$ & $0.40\pm0.12$ & $0.41\pm0.11$  \\ 
 ${\cal F}_{D^*}$ ($\times 10^{-3}$)   
                          & $0.25\pm0.06$ & $0.23\pm0.06$ & $0.02\pm0.01$ & 0 & 0  \\ \hline
\end{tabular}
\caption[]{Number of fitted $D\pi\ell^-$ candidates;
reconstruction times selection efficiencies of the $D\ell^-$ and \pid (or \kad) from \bdoul decays;
correction factors introduced in equation~(2).
Errors are statistical only (except for $f_{\tau,c\rightarrow\ell}$ and ${\cal F}_{H}$).
Note that most of the \ddpi candidates have been removed from the selected \dz sample;
the \dz sample also includes $D^{*0}\rightarrow D^0 \pi^0$ (or $\gamma$) decays;
the \dplus sample also includes $D^{*+}\rightarrow D^+ \pi^0$ (or $\gamma$) decays.
} 
\label{tab:piyield}
\end{center}
\end{table}

\subsection{$D^{**}\ell^-$ efficiency}
 
 The semileptonic branching fraction of a $b$ quark into $D\pi$ final state was then measured as follows:
\begin{eqnarray}
      {\mathrm BR}(b \rightarrow D\pi\ell^-X) 
  =  \dfrac{\epsilon_Z}{N_Z} \; \dfrac{1}{2 R_b} \; 
     \dfrac{N(D\pi\ell^-)}{2 \; \epsilon_{D\ell} \; \epsilon_{**}} \; 
     \dfrac{f_{\tau_B}}{f_{\mathrm cor}} \; \dfrac{1-f_{\tau,c\rightarrow\ell}}{{\mathrm BR}_{D}} 
     \; - \; {\cal F}_{H} \; - \; {\cal F}_{D^*}
\end{eqnarray}
where 
$N_Z$ and $\epsilon_Z$ are defined in Section~3,
$R_{b}=0.2166\pm0.0007$ is the $Z$ hadronic decay rate into \bb events~\cite{ref:LEPEWWG};
the branching fractions, BR$_{D}$, in the three decay modes
BR$(D^0 \rightarrow K^-\pi^+)=0.0385\pm0.0009$,
BR$(D^+ \rightarrow K^-\pi^+\pi^+)=0.090\pm0.006$ and
BR$(D^{*+} \rightarrow D^0\pi_*^+)=0.683\pm0.014$ are used~\cite{ref:PDG}.
The efficiencies to reconstruct and select the $D\ell^-$ and \pid (or \kad) candidates from \bdoul decays,
denoted $\epsilon_{D\ell}$ and $\epsilon_{**}$ respectively, are indicated in Table~\ref{tab:piyield}.
They were obtained from the simulation and corrected by the factors 
$f_{\mathrm cor}$ and $f_{\tau_B}$ which are described below.
The correction factors $f_{\tau,c\rightarrow\ell}$ 
and ${\cal F}_{H}$ account for the ``$\tau,c\rightarrow\ell$" and ``hadronic" backgrounds introduced in Section~5.2;
${\cal F}_{D^*}$ is the background due to residual $D^{*+}\pi^-\ell^-$   
which applies to the ``wrong" sign $D^0\pi^-\ell^-$ and $D^0K^-\ell^-$ samples only.

\subsubsection{Efficiency correction}

The correction to the reconstruction and selection efficiency is expressed as \\
$f_{\mathrm cor}=f_{K\pi} \; f_{M_D} \; f_{\mathrm VD} \; f_{D{\mathrm vtx}} \; f_{K{\mathrm id}} \; f_{\pi{\mathrm id}}$:
\begin{enumerate}
\item [$\bullet$] 
In the $D^0\ell$ and $D^{*+}\ell$ samples, 
only the $K\pi$ decay channel was used to estimate the $\epsilon_{D\ell}$ efficiency.
For these samples,
$f_{K\pi} = N(D\ell)/N(D_{K\pi}\ell)$ where \\
$N(D\ell)=N(D\ell^-)-N(D\ell^+)$ is the difference between the total number of $D\ell$ candidates quoted in 
Table~\ref{tab:dyield} and $N(D_{K\pi}\ell)$ is the same difference computed in the $K\pi$ decay mode only of the $D^0$.
In the $D^+\ell$ sample, $f_{K\pi} = 1$.

\item [$\bullet$] 
Due to the $D$ mass ranges required in Table~\ref{tab:dyield}, $f_{M_D}$ accounts for the mass width differences
observed in real data and simulation.

\item [$\bullet$] 
A large sample of \dskpi reconstructed inside $b$-tagged jets was used in order to estimate 
selection efficiencies related to the detector response:
the Vertex Detector information which was required for all channels ($f_{\mathrm VD}$ factor),
the vertex quality cuts for the \dplus sample ($f_{D{\mathrm vtx}}$),
and the kaon identification for the \dz and \dplus samples ($f_{K{\mathrm id}}$).
For the study of the $K^-\pi^+\pi^+$ vertex quality in the \dplus sample,
a three tracks  $K^-\pi^+\pi_*^+$ vertex was also fitted in the dedicated \ds sample 
and similar cuts were applied.

\item [$\bullet$] 
The kaon rejection (or identification) requirement of the \pid (\kad) candidates 
was also checked on the same dedicated \ds sample and a correction factor $f_{\pi{\mathrm id}}$ was inferred.
\end{enumerate}

\subsubsection{$B$ lifetime correction}

 The difference between the known values of the $B$ mesons' mean lifetimes
($\tau_{B^+} = 1.65\pm0.04$~ps, $\tau_{B^0} = 1.56\pm0.04$~ps~\cite{ref:Bosc},
$\tau_{B_s}/\tau_{B^0} \sim 0.99-1.01$~\cite{ref:Neubigi})
and that used in the Monte Carlo simulation ($\tau_B^{\mathrm MC}=1.6$~ps)
has two consequences:
\begin{enumerate}
\item [$\bullet$] It affects the decay length selection efficiencies described in Section~4.
But as these selections were applied to the decay lengths divided by their errors, 
the relative correction to $\epsilon_{D\ell}$ was found to be of about $\pm0.2\%$ only.
It was thus included in the following $f_{\tau_B}$ factor.

\item [$\bullet$] 
It also affects the shape of the impact parameter distribution of simulated \pid candidates
which is used to fit the amount of $D\pi\ell^-$ candidates in real data.
The distribution shown in Figure~\ref{fig:mcimpa}a was thus recomputed 
by weighting each simulated event and by using 
the $B^0$ mean lifetime for ``right" sign $D^0h^+\ell^-$ and ``wrong" sign $D^{(*)+}\pi^+\ell^-$ candidates,
or the $B^+$ mean lifetime for ``right" sign $D^{(*)+}\pi^-\ell^-$ and ``wrong" sign $D^0h^-\ell^-$. 
These new \pid impact parameter shapes were used to fit 
the real data distributions shown in Figures~\ref{fig:imptot}-\ref{fig:impaws}.
The difference between the number of fitted $D^{(*)} \pi_{**} \ell X$ candidates
observed with and without the weighting procedure is described by the
correction factor, $f_{\tau_B}$, given in Table~\ref{tab:piyield}.
\end{enumerate}

\subsubsection{Physical background correction}

 The physical background contributions are determined in the following way:
\begin{enumerate}
\item [$\bullet$] 
According to the simulation, still $(7.6\pm0.4\;(stat))\%$ of \ddpi remained in the \dz sample;
this value was used to determine the ${\cal F}_{D^*}$ factor.

\item [$\bullet$] 
The fraction of \btaul events is evaluated as:
\begin{eqnarray}
      f_{\tau\rightarrow\ell} = \dfrac{{\mathrm BR}(b \rightarrow \tau^-\bar{\nu}_{\tau}X) 
                        {\mathrm BR}(\tau^- \rightarrow \ell^-\bar{\nu}_{\ell}\nu_{\tau})} 
                       {{\mathrm BR}(b \rightarrow \ell^-\bar{\nu}_{\ell}X)} \; 
                 \dfrac{\epsilon_{\tau\rightarrow\ell}}{\epsilon_{\ell}} 
                               = 0.0075 \pm 0.0020
\end{eqnarray}
where BR$(b \rightarrow \tau^-\bar{\nu}_{\tau}X)=(2.6\pm0.4)\%$,
BR$(\tau^- \rightarrow \ell^-\bar{\nu}_{\ell}\nu_{\tau})=(17.64\pm0.06)\%$ and
BR$(b \rightarrow \ell^-\bar{\nu}_{\ell}X)=(10.99\pm0.23)\%$~\cite{ref:PDG};
$\epsilon_{\tau\rightarrow\ell}/\epsilon_{\ell}=0.18\pm0.04 \;(stat)$ is the ratio of the lepton selection efficiencies
in $b \rightarrow \tau^-\bar{\nu}_{\tau}X$ and $b \rightarrow \ell^-\bar{\nu}_{\ell}X$ simulated events.

 The fraction of \bcbarl events is evaluated as:
\begin{eqnarray}
      f_{\overline{c}\rightarrow\ell} = \dfrac{{\mathrm BR}(b \rightarrow \overline{c} \rightarrow \ell)}
                       {{\mathrm BR}(b \rightarrow \ell^-\bar{\nu}_{\ell}X)} \; 
                 \dfrac{\epsilon_{c\rightarrow\ell}}{\epsilon_{\ell}} 
                              = 0.047 \pm 0.012
\end{eqnarray}
where BR$(b \rightarrow \overline{c} \rightarrow \ell))=(1.6\pm0.4)\%$~\cite{ref:LEPHF};
$\epsilon_{c\rightarrow\ell}/\epsilon_{\ell}=0.32\pm0.02 \;(stat)$ is the ratio of the lepton selection efficiencies
in \bcbarl and $b \rightarrow \ell^-\bar{\nu}_{\ell}X$ simulated events.

 The fraction of \bcl events is obtained as previously,
but the probability, $P_{W \rightarrow D}$,
for the virtual $W^-$ to decay into a $\overline{D}^0$ or $D^-$ meson has to be taken into account:
\begin{eqnarray}
      f_{c\rightarrow\ell} = \dfrac{{\mathrm BR}(b \rightarrow c \rightarrow \overline{\ell})}
                                   {{\mathrm BR}(b \rightarrow \ell^-\bar{\nu}_{\ell}X)} \; 
                 P_{W \rightarrow D} \;
                 \dfrac{\epsilon_{c\rightarrow\ell}}{\epsilon_{\ell}} 
                              = 0.020 \pm 0.005
\end{eqnarray}
where 
BR$(b \rightarrow c \rightarrow \overline{\ell})=(7.8\pm0.6)\%$~\cite{ref:PDG} and 
$P_{W \rightarrow D}=(9.0\pm1.9)\%$~\cite{ref:LEPHF}.
Finally the fraction of \ddou not issued from a direct semileptonic $b$ decay is evaluated to be:
\begin{eqnarray}
f_{\tau,c\rightarrow\ell} = f_{\tau\rightarrow\ell} + f_{\overline{c}\rightarrow\ell} + f_{c\rightarrow\ell} 
                          = 0.075 \pm 0.017 \pm 0.025 \; .
\end{eqnarray}
In the first error, which is the sum of the uncertainties quoted in equations~(3-5),
the errors on $f_{\overline{c}\rightarrow\ell}$ and $f_{c\rightarrow\ell}$ have been added linearly
because $P_{W \rightarrow D}$ was used in reference~\cite{ref:LEPHF}
to evaluate BR$(b \rightarrow \overline{c} \rightarrow \ell)$.
The second error in equation~(6) is an estimate of the uncertainty due to possible phase space or QCD corrections 
between the $b \rightarrow \ell$ and the $b \rightarrow \tau,c \rightarrow \ell$ decay channels
with a $D^{**}$ in the final state. 

\item [$\bullet$] 
 The ``hadronic" background is evaluated as: 
\begin{eqnarray}
{\cal F}_{H} = {{\mathrm BR}(b \rightarrow D X)} \; (1-P_{W \rightarrow D}) \;
{\mathrm BR}(b \rightarrow (c \;{\mathrm or}\; \overline{c}) \rightarrow \ell^{\pm}) \;
                 \dfrac{\epsilon_{b\rightarrow D H \ell X}}{\epsilon_{D\ell}\epsilon_{**}} 
\end{eqnarray}
with
\begin{eqnarray}
{\mathrm BR}(b \rightarrow (c \;{\mathrm or}\; \overline{c}) \rightarrow \ell^{\pm}) &=&
       {\mathrm BR}(b \rightarrow \overline{c} \rightarrow \ell) +
       {\mathrm BR}(b \rightarrow c \rightarrow \overline{\ell}) \; P_{W \rightarrow D} \nonumber \\ 
&=& (2.30\pm0.56)\% \; ,
\end{eqnarray}
BR$(b \rightarrow D^0 X)=(60.1\pm3.2)\%$~\cite{ref:PDG},
BR$(b \rightarrow D^+ X)=(23.0\pm2.1)\%$ and BR$(b \rightarrow D^{*+} X)=(23.1\pm1.3)\%$~\cite{ref:DelphiRC}; 
the difference 
BR$(b \rightarrow D^0 X) - $BR$(b \rightarrow D^{*+} X)$BR$(D^{*+} \rightarrow D^0\pi^+)=(44.3\pm3.3)\%$ 
is used for the $D^0\ell^-$ analysis, where the \ds were rejected. 
In the $D\pi\ell^-$ analyses,
$\epsilon_{b\rightarrow D H \ell X}/(\epsilon_{D\ell}\epsilon_{**})=0.084\pm0.010 \;(stat)$
was determined from the simulation as the ratio of selection efficiencies 
between hadrons from charm decay in \bcbarl events, and genuine \pid in \bdoul decay;
in the $D^0K\ell^-$ analysis, this ratio was estimated to be $0.030\pm0.006 \;(stat)$.
The resulting ${\cal F}_{H}$ values are reported in Table~\ref{tab:piyield}.
\end{enumerate}

\subsection{Systematics}
 
 The systematics are detailed in Table~\ref{tab:syst}.
As a cross-check of the procedure, the same analysis was repeated on simulated \qq and \bb samples:
\begin{enumerate}
\item [$\bullet$] $1998\pm107$ ($1017\pm73$, $870\pm62$) \ddou candidates were fitted in the 
``right" sign $D^0\pi^+\ell^-$ ($D^+\pi^-\ell^-$, $D^{*+}\pi^-\ell^-$) samples
whereas 1934 (1106, 879) \ddou were expected;
\item [$\bullet$] $396\pm67$ ($219\pm52$, $60\pm38$) \ddou candidates were fitted in the 
``wrong" sign $D^0\pi^+\ell^-$ ($D^+\pi^-\ell^-$, $D^{*+}\pi^-\ell^-$) samples
whereas 333 (235, 62) \ddou were expected.
\end{enumerate}
A good agreement was thus obtained in the simulation between the fitted and expected \pid contributions,
the related statistical error being used to estimate the systematic uncertainty due to
the subtraction of the fake $D$ and fake lepton backgrounds.
The remaining statistical error of the Monte Carlo simulation is due to the limited number of generated
$\bar{B} \rightarrow D\pi\ell^-\bar{\nu}_{\ell}X$ events.

Following the detailed study of reference~\cite{ref:Bmulti},
a $\pm0.3\%$ uncertainty is assigned to the reconstruction efficiency of each charged particle.

 The uncertainty on the impact parameter resolution has two sources:
\begin{enumerate}
\item [$\bullet$] Impact parameter relative to the primary interaction vertex: \\
the uncertainty on the parameters $a$ and $b$ of equation~(1) affects the impact parameter distributions
of Figure~\ref{fig:mcimpa} and thus the result of the fit to the real data;
the corresponding relative systematic error is estimated to be at most of $\pm1\%$.
\item [$\bullet$] Selection of the impact parameter of the \pid candidate relative to the $D(\pi_*)\ell^-$ vertex: \\
this impact parameter was required to be smaller than 100~$\mu$m, which allowed the selection of about 82\% of 
genuine \pid candidates.
A variation of $\pm10\%$ of the impact parameter resolution modified the relative efficiency by about $\pm2.5\%$,
the magnitude depending on the considered $D$ decay channel.
A similar uncertainty was also inferred by comparing, in $D^{*+}\ell^-$ real data and simulation, the effect of a cut on the
impact parameter of the lepton relative to the $D^0\pi_*\ell^-$ vertex.
\end{enumerate}
The overall systematic uncertainty due to the impact parameter resolution is given in Table~\ref{tab:syst}.

 The uncertainty due to the \pid momentum spectrum was evaluated by comparing the \pid selection efficiencies in 
simulated $D_J \rightarrow D\pi$ and $D\pi\pi$ decays: a relative difference of $\pm1\%$ was observed.
\begin{table}[bth]
\begin{center}
\begin{tabular}{|c||c|c|c|c|c|} \hline
 Error source             &  $D^0h\ell^-$ & $D^0\pi\ell^-$ & $D^0K\ell^-$ & $D^+\pi\ell^-$ & $D^{*+}\pi\ell^-$ \\ \hline\hline
 $b \rightarrow \ell$ decay model~\cite{ref:LEPEWWG}
                          &  $\pm 1.2$  &  $\pm 1.2$  &  $\pm 1.2$  &  $\pm 1.2$  &  $\pm 1.2$           \\
 $\tau_B$~\cite{ref:Bosc}
                          &  $\pm 1.9$  &  $\pm 1.8$  &  $\pm 2.2$  &  $\pm 1.6$  &  $\pm 1.2$           \\
 \pid momentum spectrum
                          &  $\pm 1.0$  &  $\pm 1.0$  &  $\pm 1.0$  &  $\pm 1.0$  &  $\pm 1.0$           \\
 $\tau,c \rightarrow \ell$ background
                          &  $\pm 3.5$  &  $\pm 3.5$  &  $\pm 4.2$  &  $\pm 3.5$  &  $\pm 3.5$           \\
 ``hadronic" background
                          &  $\pm 2.5$  &  $\pm 2.1$  &  $\pm 9.2$  &  $\pm 2.4$  &  $\pm 2.3$           \\  
 BR$(D^0 \rightarrow K^-\pi^+)$~\cite{ref:PDG}      
                          &  $\pm 2.3$  &  $\pm 2.3$  &  $\pm 2.3$  &     --      &  $\pm 2.3$           \\  
 BR$(D^+ \rightarrow K^-\pi^+\pi^+)$~\cite{ref:PDG} & \multicolumn{3}{c|}{--}     &  $\pm 6.7$  &  --    \\ 
 BR$(D^{*+} \rightarrow D^0\pi^+)$~\cite{ref:PDG}   & \multicolumn{3}{c|}{--}     &  --   &  $\pm 2.0$   \\ \hline\hline
 fake $D\ell$ backgrounds &  $\pm 5.4$  &  $\pm 5.4$  &  $\pm 5.4$  &  $\pm 7.2$  &  $\pm 7.1$  \\ 
 MC statistics            &  $\pm 2.0$  &  $\pm 2.0$  &  $\pm 2.3$  &  $\pm 2.6$  &  $\pm 1.7$  \\ 
 track reconstruction     &  $\pm 1.3$  &  $\pm 1.3$  &  $\pm 1.5$  &  $\pm 1.6$  &  $\pm 1.6$  \\
 impact parameter resolution
                          &  $\pm 3.2$  &  $\pm 3.1$  &  $\pm 3.7$  &  $\pm 3.7$  &  $\pm 2.8$  \\
 mass resolution          &  $\pm 1.2$  &  $\pm 1.1$  &  $\pm 1.4$  &  $\pm 1.1$  &  $\pm 1.1$  \\ 
 VD requirement           &  $\pm 3.3$  &  $\pm 3.2$  &  $\pm 3.9$  &  $\pm 3.2$  &  $\pm 3.3$  \\ 
 $D$ vertex selection     & \multicolumn{3}{c|}{--}                 &  $\pm 3.3$  &     --      \\ 
 $K$ (from $D$) identification  
                          &  $\pm 2.6$  &  $\pm 2.6$  &  $\pm 3.1$  &  $\pm 2.6$  &     --      \\
 \pid (or \kad) identification  
                          &      --     &  $\pm 2.1$  & $\pm 18.1$  &  $\pm 1.2$  &  $\pm 1.2$  \\ 
 lepton identification    &  $\pm 2.0$  &  $\pm 1.9$  &  $\pm 2.3$  &  $\pm 2.0$  &  $\pm 2.0$  \\ \hline\hline
 Total                    &  $\pm 9.9$  &  $\pm 9.9$  & $\pm 22.9$  & $\pm 13.3$  & $\pm 10.6$  \\ \hline 
\end{tabular}
\caption[]{Relative systematic uncertainties ($\%$) on the $b$ semileptonic branching fractions
into $D^{**}\ell^-\bar{\nu}_{\ell}$ final states (``right" sign only).} 
\label{tab:syst}
\end{center}
\end{table}

\subsection{Results}
 
 From the previous study, the $b$ semileptonic branching fraction can be computed in each $D\pi\ell^-$
or $D^0K\ell^-$ final state. 
The corresponding results are reported in Table~\ref{tab:br} which includes the statistical and systematic errors.
The ``right" sign values are in agreement with those measured by the ALEPH collaboration~\cite{ref:ALEPH},
except for the $D^0\pi^+\ell^-X$ channel where the DELPHI result is two standard deviations larger.

 The ``wrong" sign results are at less than 2 standard deviations from zero,
thus $D\pi\pi$ final states will be neglected in the following.
The $D^0K\ell^-$ production rate is also found to be compatible with zero.
Thus only $D\pi\ell^-$ final states will be considered in the following.
 As a further cross-check, Tables~\ref{tab:check2}-\ref{tab:check3} 
present the $b$ semileptonic branching fraction measurement for electrons and muons separately
and for the various $D$ decay channels.

 Using the production fraction BR$(b \rightarrow \bar{B}^0)=$BR$(b \rightarrow B^-)=0.395\pm0.014$~\cite{ref:Bosc},
the following branching fractions are measured:
\begin{eqnarray*}
{\mathrm BR}(\bar{B}^0 \rightarrow D^0 \pi^+ \ell^- \bar{\nu_{\ell}}) +
{\mathrm BR}(\bar{B}^0 \rightarrow D^{*0} \pi^+ \ell^- \bar{\nu_{\ell}}) 
                      &=& (2.70\pm0.64\;(stat)\pm0.28\;(syst))\%            \\
{\mathrm BR}(B^- \rightarrow D^+ \pi^- \ell^- \bar{\nu_{\ell}}) + 
{\mathrm BR}(B^- \rightarrow D^{*+} \pi^- \ell^- \bar{\nu_{\ell}}) 
                      &=& (2.08\pm0.47\;(stat)\pm0.20\;(syst))\% \; .
\end{eqnarray*}
\begin{table}[bth]
\begin{center}
\begin{tabular}{|c|c|c|c|c|} \hline
 \multicolumn{5}{|l|}{BR$(b \rightarrow D^{**} \ell^- \bar{\nu}_{\ell})$ ($\times 10^{-3}$)} \\
   $D^0h\ell^-X$ & $D^0\pi\ell^-X$ & $D^0K\ell^-X$ & $D^+\pi\ell^-X$ & $D^{*+}\pi\ell^-X$  \\ \hline\hline
 \multicolumn{5}{|l|}{DELPHI ``right" sign} \\
 $11.6\pm2.4\pm1.1$ & $10.7\pm2.5\pm1.1$ & $~1.0\pm1.1\pm0.2$ & $4.9\pm1.8\pm0.7$ & $4.8\pm0.9\pm0.5$ \\ \hline 
 \multicolumn{5}{|l|}{DELPHI ``wrong" sign} \\
  $~1.9\pm1.4\pm0.4$ & $~2.3\pm1.5\pm0.4$ & $-0.4\pm0.6\pm0.1$ & $2.6\pm1.5\pm0.4$ & $0.6\pm0.7\pm0.2$ \\ 
\hline \hline 
 \multicolumn{5}{|l|}{ALEPH ``right" sign} \\
   -- & $~4.7\pm1.3\pm1.0$ &  $~2.6\pm1.2\pm0.8$ & $3.0\pm0.7\pm0.5$ & $4.7\pm0.8\pm0.6$ \\ \hline
\end{tabular}
\caption[]{Semileptonic branching fractions BR$(b \rightarrow D^{**} \ell^- \bar{\nu}_{\ell})$ 
measured in DELPHI for each $D\pi\ell^-$ or $D^0K\ell^-$ final state. 
Similar results obtained in ALEPH are also presented~\cite{ref:ALEPH}.
The first errors are statistical and the second systematic.
Note that $D^0$s from the \ddpi decay mode are removed from the $D^0\ell^-$ results,
which still include $D^0$s from $D^{*0}\rightarrow D^0 \pi^0$ (or $\gamma$) decays;
results on \dplus also include $D^{*+}\rightarrow D^+ \pi^0$ (or $\gamma$) decays.
} 
\label{tab:br}
\end{center}
\end{table}
\begin{table}[bth]
\begin{center}
\begin{tabular}{|c|c|c|c|} \hline
 BR$(b \rightarrow D^{**} \ell^- \bar{\nu}_{\ell})$ ($\times 10^{-3}$)
           & $D^0h^+\ell^-X$ & $D^+\pi^-\ell^-X$ & $D^{*+}\pi^-\ell^-X$  \\ \hline\hline
 $e$       &  $~8.8\pm3.0$  &    $5.3\pm2.3$   &    $5.1\pm1.2$       \\ 
 $\mu$     &  $14.6\pm3.4$  &    $4.5\pm2.5$   &    $4.4\pm1.2$       \\ \hline
 Average   &  $11.6\pm2.4$  &    $4.9\pm1.8$   &    $4.8\pm0.9$       \\ \hline
\end{tabular}
\caption[]{Semileptonic branching fraction for electrons and muons separately.
Errors are statistical only.} 
\label{tab:check2}
\end{center}
\end{table}
\begin{table}[bth]
\begin{center}
\begin{tabular}{|c|c|c|} \hline
 BR$(b \rightarrow D^{**} \ell^- \bar{\nu}_{\ell})$ ($\times 10^{-3}$)
               & $D^0h^+\ell^-X$ & $D^{*+}\pi^-\ell^-X$  \\ \hline\hline
 $K\pi$        &  $13.6\pm2.6$  &    $5.5\pm1.5$       \\ 
 $K\pi\pi\pi$  &  $~9.8\pm4.1$  &    $5.7\pm1.8$       \\ 
 $K\pi\pi^0$   &                &    $3.9\pm1.3$       \\ \hline
 Average       &  $11.6\pm2.4$  &    $4.8\pm0.9$       \\ \hline
\end{tabular}
\caption[]{Semileptonic branching fraction for the different $D$ decay channels.
Errors are statistical only.} 
\label{tab:check3}
\end{center}
\end{table}

According to isospin conservation rules
and assuming that only $D\pi$ and $D^*\pi$ final states contribute,
the ratios between final states involving charged and neutral pions 
are predicted to be:
\begin{eqnarray}
 \dfrac{D^0\pi^+ + D^{*0}\pi^+}{D^+\pi^0 + D^{*+}\pi^0} = 
 \dfrac{D^+\pi^- + D^{*+}\pi^-}{D^0\pi^0 + D^{*0}\pi^0} = 2 \; ,
\end{eqnarray}
allowing the following branching fractions to be inferred:
\begin{eqnarray}
{\mathrm BR}(\bar{B}^0 \rightarrow D \pi \ell^- \bar{\nu_{\ell}}) +
{\mathrm BR}(\bar{B}^0 \rightarrow D^{*} \pi \ell^- \bar{\nu_{\ell}}) 
                      = (4.05\pm0.96\;(stat)\pm0.42\;(syst))\%            \\
{\mathrm BR}(B^- \rightarrow D \pi \ell^- \bar{\nu_{\ell}}) + 
{\mathrm BR}(B^- \rightarrow D^{*} \pi \ell^- \bar{\nu_{\ell}}) 
                      = (3.12\pm0.71\;(stat)\pm0.30\;(syst))\%
\end{eqnarray}
These values are in good agreement.
Neglecting $D\pi\pi$ final states and using as a constraint the equality of equations~(10) and~(11),
the overall $B$ meson semileptonic branching fraction into any $D^{(*)}\pi$ final state can be obtained:
\begin{eqnarray*}
{\mathrm BR}(\bar{B} \rightarrow D \pi \ell^- \bar{\nu_{\ell}}) + 
{\mathrm BR}(\bar{B} \rightarrow D^{*} \pi \ell^- \bar{\nu_{\ell}}) 
                      = (3.40\pm0.52\;(stat)\pm0.32\;(syst))\%  \; .
\end{eqnarray*}

Assuming that the isospin invariance used in equation~(9) applies also to 
$D\pi$ and $D^*\pi$ separately, the following branching fractions are also inferred:
\begin{eqnarray*}
{\mathrm BR}(\bar{B} \rightarrow D \pi \ell^- \bar{\nu_{\ell}})     &=& (1.54 \pm 0.61 \;(stat+syst))\% \\
{\mathrm BR}(\bar{B} \rightarrow D^{*} \pi \ell^- \bar{\nu_{\ell}}) &=& (1.86 \pm 0.38 \;(stat+syst))\% 
\end{eqnarray*}
with a correlation coefficient of -0.33 between the results.

\section{Overall $b$ decay rate into $D \ell^- \bar{\nu}_{\ell} X$ final states} 

 In this section, a measurement of the semileptonic branching fractions BR$(b \rightarrow D \ell^- X)$,
where $D$ stands for $D^0$, \dplus or $D^{*+}$, is presented.
The method used is similar to that described in Section~5.4:
\begin{eqnarray}
      {\mathrm BR}(b \rightarrow D\ell^-X) 
  =  \dfrac{\epsilon_Z}{N_Z} \; \dfrac{1}{2 R_b} \; 
     \dfrac{N(D_{K\pi}\ell)}{2 \; \epsilon_{D\ell} \; \eta_{D\ell}} \; 
     \dfrac{1}{f^{'}_{\mathrm cor}} \; \dfrac{1-f_{\tau,c\rightarrow\ell}}{{\mathrm BR}_D} 
     \; - \; {\cal F}^{'}_{D^*} \; + \; {\cal F}_{Dh\ell}
\end{eqnarray}
where 
$N(D_{K\pi}\ell)=N(D\ell^-)-N(D\ell^+)$ is the difference between the total number of $D\ell$ candidates quoted in 
Table~\ref{tab:dyield}, using only the $K\pi$ ($K\pi\pi$, $K\pi\pi_*$) decay mode in the $D^0$ ($D^+$, $D^{*+}$) analyses;
$f^{'}_{\mathrm cor}=f_{M_D} \; f_{\mathrm VD} \; f_{D{\mathrm vtx}} \; f_{K{\mathrm id}}$;
${\cal F}^{'}_{D^*}$ is only used for the \dz sample where a fraction of $(7.6\pm0.4\;(stat))\%$ of \ddpi decays is included.

 According to the simulation, the reconstruction efficiency of $D\ell$ final states depends slightly 
on whether or not the $D$ meson originates from a $D^{**}$.
In the absence of $D^{**}$, the reconstruction times selection efficiency, $\epsilon_{D\ell}$, 
of Table~\ref{tab:piyield}
has to be multiplied by the factor $\eta_{D\ell} = 1.08\pm0.02$ ($1.13\pm0.02$, $1.07\pm0.02$)
for a $D^0$ ($D^+$, $D^{*+}$) final state.
Thus the 
observed production fraction of $D\pi\ell^-$ and $D^0K\ell^-$ final states (denoted $Dh\ell^-$)
has to be taken into account in equation~(12) and the following factor is introduced:
\begin{eqnarray}
  {\cal F}_{Dh\ell} = \dfrac{\eta_{D\ell}-1}{\eta_{D\ell}} \; {\mathrm BR}(b \rightarrow Dh\ell^-X) \; .
\end{eqnarray}

 The overall $b$ semileptonic branching fractions are thus measured to be:
\begin{eqnarray}
{\mathrm BR}(b \rightarrow D^0 \ell^- \bar{\nu}_{\ell} X)    = 
 (7.04\pm0.34\;(stat)\pm0.36\;(syst.exp)\pm0.17\;({\mathrm BR}_D))\% \nonumber \\
{\mathrm BR}(b \rightarrow D^+ \ell^- \bar{\nu}_{\ell} X)    = 
 (2.72\pm0.19\;(stat)\pm0.16\;(syst.exp)\pm0.18\;({\mathrm BR}_D))\% \\
{\mathrm BR}(b \rightarrow D^{*+} \ell^- \bar{\nu}_{\ell} X) = 
 (2.75\pm0.17\;(stat)\pm0.13\;(syst.exp)\pm0.09\;({\mathrm BR}_D))\% \nonumber
\end{eqnarray}
where the $D^0 \ell^-$ result includes also \dz coming from $D^{*0}$ and 
also (contrarily to Section~5) \ddpi decays,
the $D^+ \ell^-$ result includes also \dplus coming from $D^{*+}\rightarrow D^+ \pi^0$ (or $\gamma$) decays
and $X$ means ``anything" (possibly a hadron coming from a $D^{**}$).
These results are compared in Table~\ref{tab:DELPHIOPAL} with those measured by the OPAL collaboration~\cite{ref:OPAL}:
the $D^0$ and $D^{*+}$ values are in agreement whereas the $D^+$ results
present a difference of two standard deviations.
The systematics are detailed in Table~\ref{tab:systdlx}.

 The relative yield of $D^{*+}\pi\ell^-\bar{\nu}_{\ell}X$ over all $D^{*+}\ell^-\bar{\nu}_{\ell}X$ 
is a contribution to the systematic uncertainty
of various measurements, particularly of $V_{cb}$~\cite{ref:LEPHFS,ref:VCB,ref:VCBAO}. 
From Table~\ref{tab:br} and {\mbox equations}~(9) and~(14), the following ratio is obtained:
\begin{eqnarray*}
\dfrac{{\mathrm BR}(b \rightarrow D^{*+} \pi^- \ell^- \bar{\nu}_{\ell} X) +
       {\mathrm BR}(b \rightarrow D^{*+} \pi^0 \ell^- \bar{\nu}_{\ell} X)}
{{\mathrm BR}(b \rightarrow D^{*+} \ell^- \bar{\nu}_{\ell} X)} = 0.26\pm0.05\;(stat)\pm0.02\;(syst)
\end{eqnarray*}
which significantly improves on a previous DELPHI measurement~\cite{ref:VCB}. 
\begin{table}[bth]
\begin{center}
\begin{tabular}{|c|c|c|c|} \hline
 BR$(b \rightarrow D \ell^- \bar{\nu}_{\ell} X)$ 
           &          $D^0\ell^-$        &          $D^+\ell^-$        &          $D^{*+}\ell^-$     \\
 ($\%$)    &                             &                             &                             \\ \hline\hline
 DELPHI    & $7.04\pm0.34\pm0.36\pm0.17$ & $2.72\pm0.19\pm0.16\pm0.18$ & $2.75\pm0.17\pm0.13\pm0.09$ \\ 
 OPAL      & $6.55\pm0.36\pm0.44\pm0.15$ & $2.02\pm0.22\pm0.13\pm0.14$ & $2.86\pm0.18\pm0.21\pm0.09$ \\ \hline
\end{tabular}
\caption[]{Overall semileptonic branching fractions into $D\ell^-$ final states
as measured in DELPHI and OPAL~\cite{ref:OPAL}.
The first errors are statistical, the second are experimental systematics and the last 
are due to the exclusive $D$ branching fractions, BR$_D$.} 
\label{tab:DELPHIOPAL}
\end{center}
\end{table}
\begin{table}[bth]
\begin{center}
\begin{tabular}{|c||c|c|c|} \hline
 Error source             &  $D^0\ell^-$ & $D^+\ell^-$ & $D^{*+}\ell^-$ \\ \hline\hline
 $b \rightarrow \ell$ decay model~\cite{ref:LEPEWWG}
                          &  $\pm 1.2$  &  $\pm 1.2$  &  $\pm 1.2$           \\
 $\tau_B$~\cite{ref:Bosc}
                          &  $\pm 0.2$  &  $\pm 0.2$  &  $\pm 0.2$           \\
 $\tau,c \rightarrow \ell$ background
                          &  $\pm 1.8$  &  $\pm 1.8$  &  $\pm 1.8$           \\
 BR$(D^0 \rightarrow K^-\pi^+)$~\cite{ref:PDG}      
                          &  $\pm 2.3$  &     --      &  $\pm 2.3$           \\  
 BR$(D^+ \rightarrow K^-\pi^+\pi^+)$~\cite{ref:PDG} &  --  &  $\pm 6.7$  &  --    \\ 
 BR$(D^{*+} \rightarrow D^0\pi^+)$~\cite{ref:PDG}   &  --  &  --  &  $\pm 2.0$   \\ \hline\hline
 MC statistics            &  $\pm 2.4$  &  $\pm 2.4$  &  $\pm 2.7$  \\ 
 track reconstruction     &  $\pm 0.9$  &  $\pm 1.2$  &  $\pm 1.2$  \\
 mass resolution          &  $\pm 1.1$  &  $\pm 1.0$  &  $\pm 1.0$  \\ 
 VD requirement           &  $\pm 2.0$  &  $\pm 2.0$  &  $\pm 2.0$  \\ 
 $D$ vertex selection     &      --     &  $\pm 3.1$  &     --      \\ 
 $K$ identification       &  $\pm 2.4$  &  $\pm 2.4$  &     --      \\
 lepton identification    &  $\pm 1.8$  &  $\pm 1.8$  &  $\pm 1.8$  \\ \hline\hline
 Total                    &  $\pm 5.5$  &  $\pm 9.0$  &  $\pm 5.6$  \\ \hline 
\end{tabular}
\caption[]{Relative systematic uncertainties ($\%$) on the $b$ semileptonic branching fractions
into $D\ell^-\bar{\nu}_{\ell}X$ final states.}
\label{tab:systdlx}
\end{center}
\end{table}

\section{Summary and conclusion}

Using DELPHI data recorded from 1992 to 1995,
the overall $b$ semileptonic branching fractions into $D^0$, \dplus or \ds final states have been obtained:
\begin{eqnarray*}
{\mathrm BR}(b \rightarrow D^0 \ell^- \bar{\nu}_{\ell} X)    &=& 
 (7.04\pm0.34\;(stat)\pm0.36\;(syst.exp)\pm0.17\;({\mathrm BR}_D))\% \\
{\mathrm BR}(b \rightarrow D^+ \ell^- \bar{\nu}_{\ell} X)    &=& 
 (2.72\pm0.19\;(stat)\pm0.16\;(syst.exp)\pm0.18\;({\mathrm BR}_D))\% \\
{\mathrm BR}(b \rightarrow D^{*+} \ell^- \bar{\nu}_{\ell} X) &=& 
 (2.75\pm0.17\;(stat)\pm0.13\;(syst.exp)\pm0.09\;({\mathrm BR}_D))\% 
\end{eqnarray*}
where the $D^0$ and $D^+$ results include also contributions from
$D^{*0}$ and $D^{*+}$ decays.

Evaluating the yield of charged pions from higher excited states or from non-resonant $D^{(*)}\pi$ final states,
the following branching fractions have been measured:
\begin{eqnarray*}
{\mathrm BR}(b \rightarrow D^0 \pi^+ \ell^- \bar{\nu_{\ell}} X)    &=& (10.7\pm2.5\;(stat)\pm1.1\;(syst))10^{-3} \\ 
{\mathrm BR}(b \rightarrow D^+ \pi^- \ell^- \bar{\nu_{\ell}} X)    &=& ~(4.9\pm1.8\;(stat)\pm0.7\;(syst))10^{-3} \\ 
{\mathrm BR}(b \rightarrow D^{*+} \pi^- \ell^- \bar{\nu_{\ell}} X) &=& ~(4.8\pm0.9\;(stat)\pm0.5\;(syst))10^{-3} 
\end{eqnarray*}
and
\begin{eqnarray*}
{\mathrm BR}(b \rightarrow D^0 \pi^- \ell^- \bar{\nu_{\ell}} X)    &=& ( 2.3\pm1.5\;(stat)\pm0.4\;(syst))10^{-3} \\ 
{\mathrm BR}(b \rightarrow D^+ \pi^+ \ell^- \bar{\nu_{\ell}} X)    &=& ( 2.6\pm1.5\;(stat)\pm0.4\;(syst))10^{-3} \\ 
{\mathrm BR}(b \rightarrow D^{*+} \pi^+ \ell^- \bar{\nu_{\ell}} X) &=& ( 0.6\pm0.7\;(stat)\pm0.2\;(syst))10^{-3} 
\end{eqnarray*}
where the \ddpi decay mode is not included in the BR$(b \rightarrow D^0 \pi^{\pm} \ell^- \bar{\nu_{\ell}} X)$ results.
Neglecting $D\pi\pi$ final states and assuming isospin invariance, the separated branching fractions are inferred:
\begin{eqnarray*}
{\mathrm BR}(\bar{B} \rightarrow D \pi \ell^- \bar{\nu_{\ell}})     &=& (1.54 \pm 0.61 \;(stat+syst))\% \\
{\mathrm BR}(\bar{B} \rightarrow D^{*} \pi \ell^- \bar{\nu_{\ell}}) &=& (1.86 \pm 0.38 \;(stat+syst))\% \; . 
\end{eqnarray*}
The measured overall branching fraction:
\begin{eqnarray*}
{\mathrm BR}(\bar{B} \rightarrow D \pi \ell^- \bar{\nu_{\ell}}) + 
{\mathrm BR}(\bar{B} \rightarrow D^{*} \pi \ell^- \bar{\nu_{\ell}}) 
                      = (3.40\pm0.52\;(stat)\pm0.32\;(syst))\%  
\end{eqnarray*}
is found, in good agreement with the expectation from the difference~\cite{ref:PDG}:
\begin{eqnarray*}
{\mathrm BR}(\bar{B}   \rightarrow \ell^- \bar{\nu_{\ell}} X) -
{\mathrm BR}(\bar{B}^0 \rightarrow D^+ \ell^- \bar{\nu_{\ell}}) -
{\mathrm BR}(\bar{B}^0 \rightarrow D^{*+} \ell^- \bar{\nu_{\ell}}) 
                      = (3.85\pm0.42)\%            
\end{eqnarray*}
but is larger than a previous ALEPH result of 
${\mathrm BR}(\bar{B} \rightarrow D \pi \ell^- \bar{\nu_{\ell}}) + 
{\mathrm BR}(\bar{B} \rightarrow D^{*} \pi \ell^- \bar{\nu_{\ell}}) =
(2.26\pm0.29\;(stat)\pm0.33\;(syst))\%$~\cite{ref:ALEPH}.

\subsection*{Acknowledgements}
\vskip 3 mm
 We are greatly indebted to our technical 
collaborators, to the members of the CERN-SL Division for the excellent 
performance of the LEP collider, and to the funding agencies for their
support in building and operating the DELPHI detector.\\
We acknowledge in particular the support of \\
Austrian Federal Ministry of Science and Traffics, GZ 616.364/2-III/2a/98, \\
FNRS--FWO, Belgium,  \\
FINEP, CNPq, CAPES, FUJB and FAPERJ, Brazil, \\
Czech Ministry of Industry and Trade, GA CR 202/96/0450 and GA AVCR A1010521,\\
Danish Natural Research Council, \\
Commission of the European Communities (DG XII), \\
Direction des Sciences de la Mati$\grave{\mbox{\rm e}}$re, CEA, France, \\
Bundesministerium f$\ddot{\mbox{\rm u}}$r Bildung, Wissenschaft, Forschung 
und Technologie, Germany,\\
General Secretariat for Research and Technology, Greece, \\
National Science Foundation (NWO) and Foundation for Research on Matter (FOM),
The Netherlands, \\
Norwegian Research Council,  \\
State Committee for Scientific Research, Poland, 2P03B06015, 2P03B1116 and
SPUB/P03/178/98, \\
JNICT--Junta Nacional de Investiga\c{c}\~{a}o Cient\'{\i}fica 
e Tecnol$\acute{\mbox{\rm o}}$gica, Portugal, \\
Vedecka grantova agentura MS SR, Slovakia, Nr. 95/5195/134, \\
Ministry of Science and Technology of the Republic of Slovenia, \\
CICYT, Spain, AEN96--1661 and AEN96-1681,  \\
The Swedish Natural Science Research Council,      \\
Particle Physics and Astronomy Research Council, UK, \\
Department of Energy, USA, DE--FG02--94ER40817. \\

\newpage
 
\newpage
 
\newpage
 
\newpage
 
\newpage
 
\newpage

\end{document}